\documentclass[3p]{elsarticle}
\journal{Journal of Computational Physics}

\makeatletter
\def\ps@pprintTitle{%
 \let\@oddhead\@empty
 \let\@evenhead\@empty
 \def\@oddfoot{\footnotesize\itshape
       Published in the \ifx\@journal\@empty Elsevier
       \else\@journal\fi\hfill} 
 \let\@evenfoot\@oddfoot}
\makeatother

\usepackage{amsmath}
\usepackage{animate}
\usepackage{movie15}
\usepackage{lineno,hyperref}
\modulolinenumbers[1]
\usepackage{lipsum}
\usepackage{amsfonts}
\usepackage{graphicx}
\usepackage{epstopdf}
\usepackage{algorithmic}
\usepackage{soul}
\usepackage{float}
\usepackage{multirow}
\usepackage{booktabs}
\usepackage{amsbsy}
\usepackage{amssymb}
\usepackage{amsmath} 
\usepackage{algorithmic}
\usepackage{algorithm}
\usepackage{mathtools}
\usepackage{subcaption}
\usepackage{soul}
\usepackage{color}
\usepackage{lineno}
\usepackage{ulem}

\newcommand{\cmnt}[1]{}
\newdefinition{rmk}{Remark}
\graphicspath{{Figures}/}

\newcommand{\cmt}[1]{}

\newcommand{\x}{{\bf x}}

\newcommand{\ben}{\begin{enumerate}}
\newcommand{\een}{\end{enumerate}}

\newlength\myindent
\numberwithin{equation}{section}
\begin{document}

\begin{frontmatter}

\title{A unified scalable framework for causal sweeping strategies for Physics-Informed Neural Networks (PINNs) and their temporal decompositions}
\author{Michael Penwarden$^{1,2}$, Ameya D. Jagtap$^{3}$, Shandian Zhe$^{2}$, George Em Karniadakis$^{3}$, Robert M. Kirby$^{1,2}$}
\cortext[mycorrespondingauthor]{Email addresses: 
 mpenwarden@sci.utah.edu (Michael Penwarden), ameya$\_$jagtap@brown.edu (Ameya D. Jagtap), zhe@cs.utah.edu (Shandian Zhe), george$\_$karniadakis@brown.edu (George Em Karniadakis), kirby@cs.utah.edu (Robert M. Kirby)}

\address{$^1$~Scientific Computing and Imaging Institute, University of Utah, Salt Lake City, UT 84112, USA.}
\address{$^2$~ Kahlert School of Computing, University of Utah, Salt Lake City, UT 84112, USA.}
\address{$^3$~Division of Applied Mathematics, Brown University, 182 George Street, Providence, RI 02912, USA.}
\begin{abstract}
Physics-informed neural networks (PINNs) as a means of solving partial differential equations (PDE) have garnered much attention in the Computational Science and Engineering (CS\&E) world. However, a recent topic of interest is exploring various training (i.e., optimization) challenges -- in particular, arriving at poor local minima in the optimization landscape results in a PINN approximation giving an inferior, and sometimes trivial, solution when solving forward time-dependent PDEs with no data. This problem is also found in, and in some sense more difficult, with domain decomposition strategies such as temporal decomposition using XPINNs. We furnish examples and explanations for different training challenges, their cause, and how they relate to information propagation and temporal decomposition. We then propose a new stacked-decomposition method that bridges the gap between time-marching PINNs and XPINNs. We also introduce significant computational speed-ups by using transfer learning concepts to initialize subnetworks in the domain and loss tolerance-based propagation for the subdomains. Finally, we formulate a new time-sweeping collocation point algorithm inspired by the previous PINNs causality literature, which our framework can still describe, and provides a significant computational speed-up via reduced-cost collocation point segmentation. The proposed methods form our unified framework, which overcomes training challenges in PINNs and XPINNs for time-dependent PDEs by respecting the causality in multiple forms and improving scalability by limiting the computation required per optimization iteration. Finally, we provide numerical results for these methods on baseline PDE problems for which unmodified PINNs and XPINNs struggle to train. 
\end{abstract}


\end{frontmatter}


\section{Introduction}
\label{sec:introduction}

Physics-informed neural networks (PINNs) have emerged as a popular framework for solving partial differential equations (PDEs). The most ubiquitously used PINN implementation at present is the meshless, continuous-time approach in \cite{raissi2019physics}. This approach is often selected due to its flexibility in discretization and has been shown to be successful across a wide class of application domains \cite{mathews2021uncovering,kissas2020machine,10.1371/journal.pcbi.1007575,WANG2021109914,shukla2020physics,Chen:20,10.3389/fphy.2020.00042}. However, the users community has observed that the continuous-time approach suffers from various training challenges not experienced by the discrete-time approach. In this work, we are motivated to keep as much discretization flexibility as the continuous-time approaches allow while benefiting from the properties of the discrete-time approach. We will return to this trade-off when proposing our new time-sweeping collocation point algorithm. As continuous-time PINNs have become the default form, future mentions of PINNs will refer to this approach unless explicitly stated otherwise. Comprehensive surveys of PINN methodologies can be found in \cite{cuomo2022scientific,hao2022physics}. 

Training (i.e., optimization) remains the primary challenge when using the continuous-time approach for forward problems. A significant amount of the research on PINNs revolves around improving the ease of training in some way \cite{krishnapriyan2021characterizing, wight2020solving, hu2021extended, mojgani2022lagrangian}. However, PINNs for inverse problems have shown great success on a range of applications and do not pose the same training issues as forward problems \cite{jagtap2022deep,jagtap2022physics,9664609,inverse_groundwater,chen2020physics,mishra2020estimates,thakur2023temporal}. We, therefore, focus solely on forward problems in this paper, as they are often a principal building block in solving inverse problems and also the more challenging direction when training. Information propagation drives many training challenges\cmnt{failure modes} in forward PINN problems, and in the inverse form, this is a quite different problem entirely for which forward methods might not be applicable. However, the development of forward problems for PINNs will also help drive improvements for solving inverse problems as we gain a better understanding of PINNs in general and is something to be studied in future work.

The strategies to enhance PINN training include diverse approaches such as adaptive sampling \cite{lu2021deepxde, daw2022rethinking, subramanian2022adaptive}, adaptive weighting \cite{wang2022respecting, mcclenny2020self}, adaptive activation functions \cite{jagtap2020adaptive, jagtap2022important}, additional loss terms \cite{yu2022gradient}, domain decomposition \cite{JagtapK, JAGTAP2020113028, meng2020ppinn, hu2022augmented}, and network architecture modification to obey characteristics \cite{mojgani2022lagrangian, braga2022characteristics}. A thorough summary of PINN training challenges and their proposed solutions is provided in \cite{mojgani2022lagrangian}. Recently, \cite{shin2020convergence} proposed the mathematical foundation of PINNs for linear partial differential equations, whereas \cite{mishra2022estimates} presented an estimate on the generalization error of the PINN methodology. The first comprehensive theoretical analysis of PINNs, as well as extended PINNs (XPINNs) for a prototypical nonlinear PDE, the Navier-Stokes equations have been presented in \cite{de2022error}. The optimization process of PINNs not only limits the lower bound accuracy but also causes the network to be unable to learn\cmnt{fail to learn} over the entire domain in some cases. 
Training difficulties\cmnt{failure modes} in PINNs can happen for various reasons, some of the most common being poor sampling, unequal loss term weights, or using a poor optimization scheme. 
Even with a well-tuned PINN, ``stiff" PDEs with sharp transitions \cite{wang2021understanding}, multi-scale problems \cite{wang2021eigenvector}, or highly nonlinear time-varying PDEs \cite{MATTEY2022114474} can still pose problems for the standard PINN.

Our first contribution is an experimental study and classification of training challenges \cmnt{failure modes} in PINNs and their root cause. Furthermore, we relate these training challenges\cmnt{failure modes} to information propagation during training as well as their manifestation in temporal domain decomposition strategies such as XPINNs. In doing this, we put forward a new form of training challenge\cmnt {failure mode} for XPINNs. Our next contribution is the introduction of a new unified framework to address some of these challenges\cmnt {failure modes} and highlight the current methodological gaps in PINN time-causality enforcement. PINNs and their variants are numerous and ever-increasing. Setting aside the myriad of PINN topics, time-causality considerations alone have several different approaches. A central concern facing the PINN community is the rapid development of new methods without a supporting framework between them. It is time-consuming to reimplement and retune the dozens of PINN variants (e.g., the ``alphabet" of PINN variants, cPINNs, hpPINNs, bcPINNs, etc.) and other PINN approaches for any specific problem or in use as baselines. Therefore, our approach is backward compatible with all prior methods in this regime and can easily incorporate new variants in the future. In this framework, we also bridge the gap between methods such as time-marching and XPINNs and incorporate ideas to speed up existing methods. This is done by partitioning the subdomain into collocation point sets, requiring no or small computational cost, as well as incorporating transfer learning concepts.

The paper is organized as follows: In Section \ref{sec:background}, we first summarize PINNs and related work to time-causality, which can be similarly described. We introduce a classification to these prior works and discuss the current gap in methodology. In Section \ref{sec:motivation}, we analyze different types of training challenges\cmnt {failure modes} and their relation to information propagation and decomposition. We then propose, in Section \ref{sec:methods}, a unified framework for causality-enforcing methods. Two new methods are proposed, stacked-decomposition and window-sweeping, to be used in combination with each other. These methods describe the current work covered as well as new variants. In Section \ref{sec:results}, we provide computational performance results on PDE problems with known training difficulties. We summarize and conclude our results in Section \ref{sec:conclusion}.

\section{Background}
\label{sec:background}

\subsection{Physics-Informed Neural Networks (PINNs)}
\label{ssec:pinns}

Physics-Informed Neural Networks (PINNs) were originally proposed in \cite{raissi2019physics,raissi2017physicsI,raissi2017physicsII} 
 as a neural-network-based alternative to traditional PDE discretizations. In the original PINNs work, when presented with a PDE specified over a domain $\Omega$ with boundary conditions on $\partial \Omega$ and initial conditions at $t=0$ (in the case of time-dependent PDEs), the solution is computed (i.e., the differential operator is satisfied) at a collection of collocation points. First, we rewrite our PDE system in a residual form as $\mathcal{R}(u) = \mathcal{S} - \frac{\partial}{\partial t} u - \mathcal{F}(u)$, where $\mathcal{S}$ is the source term/function and  $\mathcal{F}$ is a nonlinear operator. The PINN formulation is expressed as follows:  Given a neural network function $u_{\boldsymbol{\theta}}(\x,t)$ with specified activation functions and a weight matrix $\boldsymbol{\theta}$ denoting the degrees of freedom derived from the width and depth of the network, find  $\boldsymbol{\theta}$ that minimizes the loss function:

\begin{equation}
MSE = MSE_u + MSE_r \label{eq:mse}
\end{equation}
\noindent where
\begin{eqnarray}
MSE_u &=& \frac{1}{N_u} \sum_{i=1}^{N_u} \| u_{\boldsymbol{\theta}}(x_u^i,t_u^i) - u^i \|^2  \\
MSE_r &=& \frac{1}{N_r} \sum_{i=1}^{N_r} \| \mathcal{R}(u_{\boldsymbol{\theta}}(x_r^i,t_r^i) \|^2 \,\,\, 
\end{eqnarray}

\noindent where $\{x_u^i,t_u^i,u^i\}_{i=1}^{N_u}$ denote the initial and boundary training data on $u(\x,t)$ and $\{x^i_r,t^i_r\}_{i=1}^{N_r}$ specify the collocation points for evaluation of the collocating residual term $\mathcal{R}(u_{\boldsymbol{\theta}})$. The loss $MSE_u$ corresponds to the initial and boundary data, whereas $MSE_r$ enforces the structure imposed by the differential operator at a finite set of collocation points. For periodic boundary conditions, an exact enforcement can be used that encodes the spatial input as Fourier features \cite{dong2021method}, in which case $MSE_u$ represents only the initial condition loss. Additional terms can be added for PINN variants, such as interface terms in the case of domain-decomposition \cite{JagtapK, JAGTAP2020113028}. Often, term-wise or point-wise weights are added to Equation \ref{eq:mse} to provide improved training \cite{mcclenny2020self, wang2022and}. This loss-function minimization approach fits naturally into the traditional deep learning framework \cite{DeepLearning}. Various optimization procedures are available, including Adam \cite{kingma2014adam}, L-BFGS \cite{liu1989limited}, etc.  The procedure produces a neural network $u_{\boldsymbol{\theta}}(\x,t)$ that attempts to minimize the weak imposition of the initial and boundary conditions while satisfying the PDE residual through a balancing act. 

\subsection{Related work}
\label{ssec:related-work}

Previous works have attempted to address training issues in a variety of ways. In this section, we review relevant work that will be used as the foundation for our hypotheses and new training methods. 

\begin{figure}[H]
\centering
\includegraphics[scale=0.5]{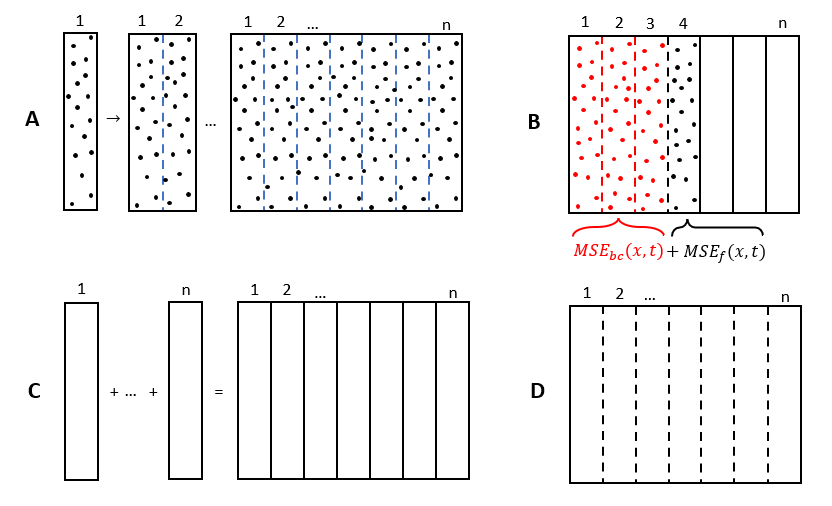}
  \caption{\small Illustrations of related models with time represented along the horizontal direction for which it progresses left to right. (A) Adaptive time-sampling. (B) Backward-compatibility. (C) Time-marching. (D) XPINNs.}
  \label{fig:related}
\end{figure}

\textbf{Adaptive time-sampling:} In \cite{wight2020solving}, a strategy is proposed that splits the domain into equally sized ``time-slabs''. For a single network, the collocation points form the sequential union of the subsets in each time-slab on which the network is continuously being trained, as seen in Figure \ref{fig:related} (A). This method is essentially a start-up procedure because it is equivalent to a standard PINN when all slab subsets have been added. This method is shown to improve training accuracy and may provide a computational speedup since only a subset of the entire spatiotemporal sampling is active in the training phase until the final slab is added. Unnecessary collocation points are expensive to add, particularly for long-time integration and higher order derivatives, because PDE residuals must be calculated for each one. 

\textbf{Time-marching:} In \cite{wight2020solving} and more recently \cite{krishnapriyan2021characterizing} and \cite{bihlo2022physics}, a training procedure is proposed in which the time-slabs are trained sequentially with the prior slab's end-time predictions used as the next initial conditions as seen in Figure \ref{fig:related} (C). Although the method's name differs between the three papers, we will refer to it here as time-marching. Since prior subnetworks stop training once a new slab is added, this enforces causality on the scale of the size of the time-slab. Internally, for each time-slab, causality is not enforced.  

\textbf{bc-PINN:} In \cite{MATTEY2022114474}, a different sequential model is proposed that, while also broken up into time-slabs, uses only one network for the entire domain. Similar to adaptive time-sampling in \cite{wight2020solving}, the difference here is that for prior time-slabs, the prediction of the converged network is taken as a data term and forms the loss with future network predictions, as seen in Figure \ref{fig:related} (B). This is termed ``backward-compatibility (bc)'' since it ensures the network does not change its prediction for prior times and is the means by which the method enforces causality. As in the time-marching scheme, this causality is enforced only on the scale of the time-slabs. Additionally, although not touched upon in the paper, this approach reduces the computational cost on a per-iteration basis since prior collocation point residuals do not need to be continually computed.

\textbf{Causal weights:} In \cite{wang2022respecting}, conforming to causality is directly confronted and put forward as a leading contributor to successful PINN training. Similar to bc-PINNs, this approach is proposed for a single network, although it is later combined with time-marching for the final numerical results on difficult chaotic problems. Unlike the previous two methods, time-slabs are not used, and instead, causality is enforced by a clever weighting mask over all collocation points. This mask is inversely exponentially proportional to the magnitude of cumulative residual losses from prior times, as shown in Equation \ref{eq:causal-weights}. One drawback is that the results are sensitive to the new causality hyperparameter $\epsilon$, so an annealing strategy for training with $\epsilon$ is used. However, this requires multiple passes over the entire domain with different $\epsilon$, significantly increasing the computational cost and not guaranteeing convergence. Despite this, its application is shown to be successful on challenging problems. 
\begin{linenomath}\begin{align}
\mathcal{L}_r \left( \boldsymbol{\theta} \right) = \frac{1}{N_t}\sum_{i=1}^{N_t} \text{exp} \left( -\epsilon \sum_{k-1}^{i-1} \mathcal{L}_r \left( t_k,\boldsymbol{\theta} \right) \right) \mathcal{L}_r \left( t_i,\boldsymbol{\theta} \right) \label{eq:causal-weights}.
\end{align}\end{linenomath}

\textbf{XPINN:} In \cite{JagtapK}, a generalized domain decomposition framework is proposed that allows for multiple subnetworks over different spatiotemporal subdomains to be stitched together and trained in parallel, as shown in Figure \ref{fig:related} (D). This method is not causal and suffers from similar training problems as standard PINNs. These problems, in some cases, become more prevalent as the interfaces and separate networks make for a more difficult optimization problem, specifically with respect to information propagation. While the idea of stitching together subdomains in time is made possible by XPINNs, time-marching and stitching together, subdomains are not mutually exclusive. Time-marching is sequential, but the networks are stitched together by the hard constraint of the final end-time prediction of the prior network used as the following initial condition. We will refer to this as the solution continuity interface condition for first-order in time problems. More precisely, it would be $MSE(u_1 - u_2)$, where $u_1$ and $u_2$ denote the predictions of the two subnetworks at the same interface locations. Alternatively, in the case of XPINNs, discontinuous enforcement by way of $MSE(u_{avg} - u_1)$ + $MSE(u_{avg} - u_2)$ where $u_{avg} = \frac{u_1+u_2}{2}$ (assuming two subdomains are intersecting along the common interface). This is extendable to second-order in time problems by adding the same forms for $u_t$ and so on for higher order in time derivative terms. While XPINNs also constrain residual continuity, this constraint is unnecessary for well-posedness when decomposing into time-slabs, such as in the prior methods discussed. In this case, the stitching between XPINNs and time-marching is the same, the difference being that the subnetworks in XPINNs are trained in parallel, and the subnetworks in time-marching are trained sequentially \cite{shukla2021parallel}. 

\subsection{Causality classification}
\label{ssec:causality-class}
Given these prior works, we seek to find a generalization between all possible methods to categorize them. We, therefore, propose the idea of hard causality and soft causality. Hard causality is a method that cannot be violated, whether continuously or discretely. Soft causality is, therefore, a method that is possible to violate; however, the network is predisposed toward obeying it in some way. This will most commonly fall under the fact that, through optimization, a network has been guided to local minima, which loosely obeys causality. A perturbation in the optimization may cause the network to find different minima, which violates this proposition, but is unlikely. We, therefore, categorize the previously described methods in Table \ref{table:causality-class}.

\begin{center}
\captionof{table}{\small A classification of PINN causality enforcement methods} 
\scalebox{0.8}{
\begin{tabular}{| c | c | c | c | c |}
\cline{2-5}
 \multicolumn{1}{c|}{} & \textbf{Soft Causality} & \textbf{Hard Causality} & \textbf{Soft + Hard Causality} & \textbf{non-Causal}\\ \hline
\textbf{Time-slab scale} & Adaptive time-sampling \cite{wight2020solving} & Time-marching \cite{wight2020solving, krishnapriyan2021characterizing, bihlo2022physics} & - & XPINN \cite{JagtapK} \\
&  & bc-PINN \cite{MATTEY2022114474} & & \\
\hline
\textbf{Sampling scale} & Causal weights \cite{wang2022respecting} & - & - & -\\
\hline
\end{tabular}}
\label{table:causality-class}
\end{center}

Notice that hard causality methods are defined only in terms of time-slabs, whereas causal weighting is a continuous form of causality. However, in the continuous case, current methods must still compute residuals for the entire domain in which they are used. There is a gap in methodology for enforcing hard causality on the sampling scale as well as for methods that combine the two. We will take inspiration from this classification to propose stacked-decomposition, which will fill the gap and form a smooth connection between a standard XPINN and time-marching, allowing for what we call causal XPINNs that overcome training issues present in their standard form. Additionally, we will use ideas from transfer learning to greatly speed up training with time-slab schemes. We will also propose a window-sweeping collocation point algorithm that will combine hard and soft causality constraints to not only speed up training by limiting the number of collocation residuals in the domain that need to be calculated, such as in adaptive time-sampling and bc-PINNs, but also enforce causality continuously. Finally, these methods can be combined to not only provide very accurate solutions, such as in  \cite{wang2022respecting}, which combines time-marching and causal weights to solve previously out-of-reach forward PINN, but also to greatly reduce the computational cost even when causality is not needed to address training challenges\cmnt {failure modes}.

\section{Training Challenges: PINNs and their temporal decompositions}
\label{sec:motivation}

\subsection{Information Propagation}
\label{ssec:motivation}

In this work, we attempt to bridge the gap between many prior works and approaches to improving PINN training. Many of these approaches are predicated on conforming to causality. Although PINNs are technically well-posed when training over the entire spatiotemporal domain represented by the set of collocation points (when properly set up), the information must still propagate from the sources of information such as initial conditions (IC) and boundary conditions (BC). We will split this discussion into two parts: first, classifying training difficulties, also called ``failure modes'' in \cite{krishnapriyan2021characterizing, daw2022rethinking, wang2021understanding}, \cmnt{failure modes} which up until now have been homogeneously grouped together; second, analyzing training difficulties\cmnt{failure modes} relating to temporal decomposition given the prior classification.

\subsubsection{Types of Training Challenges}
\label{ssec:failure_mode_types}
Let us consider two forward PDE problems. First, consider the convection problem posed in \cite{krishnapriyan2021characterizing} with enough collocation points that a standard PINN can solve it well. Second, consider the commonly used Allen-Cahn problem, which PINNs struggle to solve well without modification \cite{raissi2019physics, wight2020solving, MATTEY2022114474, wang2022respecting}. In Figure \ref{fig:failure_mode}, PINN results for three distinct types of challenges \cmnt{failures}are shown in comparison to the time-marching PINN method that results in a near approximation to the exact solution and is discussed as follows:

\begin{figure}[H]
\centering
\includegraphics[width=0.8\linewidth]{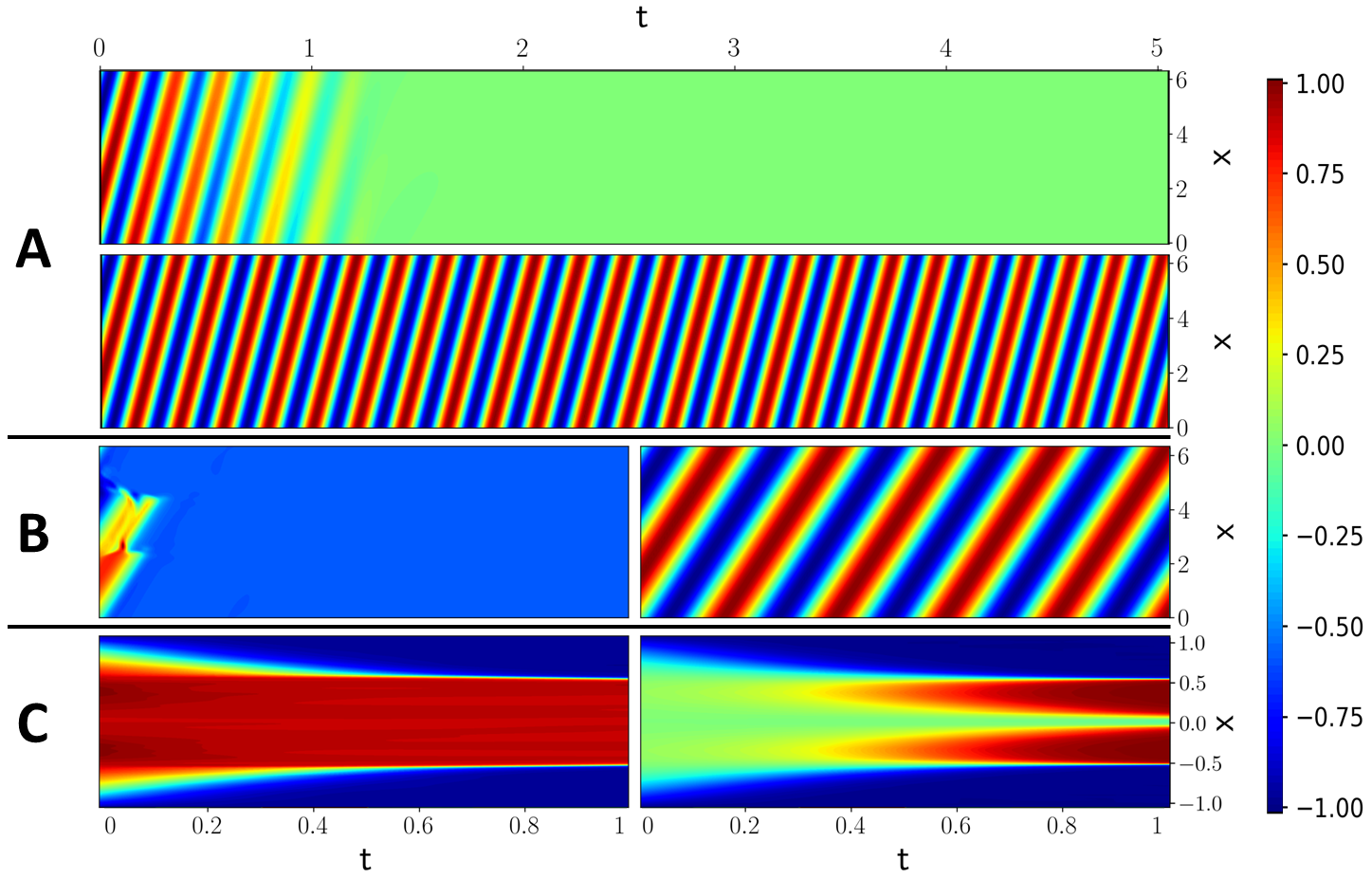}
  \caption{\small Training challenges for unmodified PINNs and the comparative accurate solution with time-marching (A) Convection problem with extended temporal domain on $T = [0, 5]$. (B) Convection problem with fewer collocation points on $T = [0, 1]$. (C) Allen-Cahn problem on $T = [0, 1]$.}\label{fig:failure_mode}
\end{figure}

\textbf{Zero-solution\cmnt{Failure}:} The zero-solution \cmnt{failure}mode is reproducible using the long-time convection problem, which extends the temporal domain to $T = [0,5]$ shown in Figure \ref{fig:failure_mode} (A). The number of residual collocation points is proportionally increased so as not to influence the result. Given periodic conditions, there is no information later in time, which results in the PINN converging to a zero-solution. This challenge occurs because the zero-solution minimizes the loss due to the PDE residual containing only derivative terms (i.e., any constant function is in the null-space of the operator). We can see that the initial condition, the only source of information, propagates in the direction of its characteristic curve. However, due to the periodic conditions, the information must travel far before being ``completed'' in  the sense that it reaches some end-point such as Dirichlet boundary conditions or the end of the time domain. When this happens, the solution can be refined. Until this happens, the propagation of information must overcome the zero-solution in the sense that the network resists the introduction of information from the initial condition. 

\begin{figure}[H]
\centering
\includegraphics[width=0.75\linewidth]{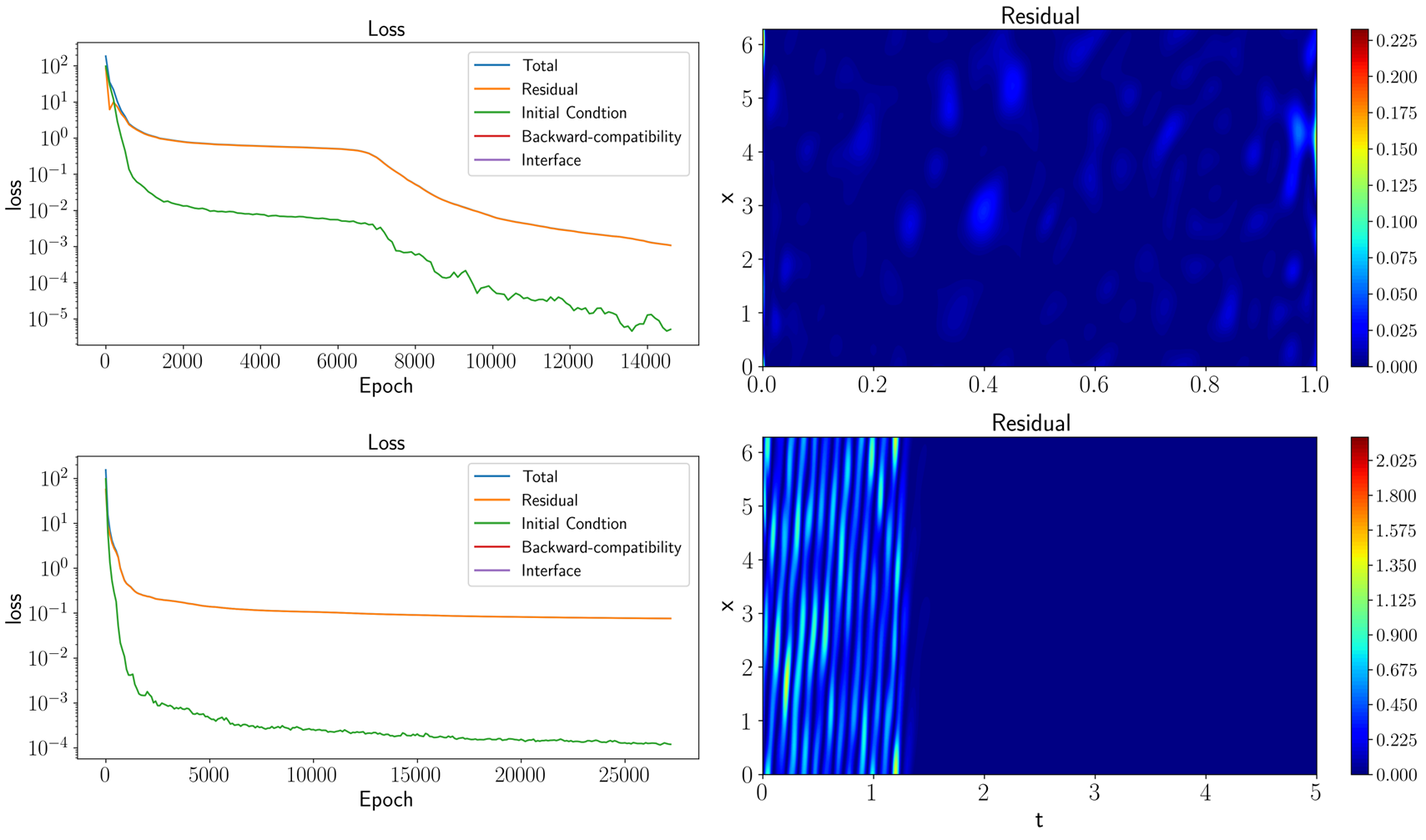}
  \caption{\small (Left) Plot of loss as a function of training epochs. (Right) The full domain PDE residual at the end of training. (Top) PINN on the convection problem with $T_{end} = 1$. (Bottom) PINN on the convection problem with $T_{end} = 5$.}\label{fig:PINN_converg}
\end{figure}

In terms of the loss landscape, the zero-solution skews it making it shallow, such that information propagates infinitesimally slowly once far enough away from the initial condition. This is shown in Figure \ref{fig:PINN_converg} where the loss and PDE residual of a trained PINN on convection $T_{end} = 1$ is shown on the top compared to $T_{end} = 5$ on the bottom. Both models are run with a termination tolerance of $10^{-7}$ measuring the change in loss per iteration. In the loss for the converged PINN, the drastic drop in the loss at around $7,500$ iterations is when the ``front" of propagation from the initial condition reaches the end of the time domain. Then, the solution refines and converges to the correct solution, minimizing the PDE residual in the domain. For a long-time problem, we can see that the residual at later times is exactly zero and therefore resists the information being propagated. Additionally, despite the variation of magnitude in the prediction, the gradients, and therefore residual, are quite uniform where the feature exists. That is to say: there is no directionality in the residual minimization at this point. Therefore, the model tries to maintain a trade-off between the loss resulting from the initial condition and its nearby collocation point residuals not being obeyed, along with the zero-solution later in time. This results in the solution petering out to zero, \cmnt{failing}never converging as it gets stuck between these two effects. 
Finally, as seen by the time-marching solution to this problem, enforcing causality can help alleviate this issue\cmnt{failure mode} since it does not allow the network to converge to the zero-solution later in time, for which the solution will not be unique until all information needed for the true solution has reached it.

\rmk{Some causal enforcement methods that still allow residual minimization later in time, such as the Lagrangian network reformulation \cite{mojgani2022lagrangian}, may improve but not fully overcome this problem \cmnt{failure mode}for an arbitrarily long enough temporal domain as the zero-solution would still be allowed.} \\

\textbf{\cmnt{Failure to}No Propagation:}\cmnt{The failure to propagate mode} This problem is reproducible by using too few residual collocation points in the convection problem shown in Figure \ref{fig:failure_mode} (B). This\cmnt{failure mode} issue is the same as the one observed in \cite{krishnapriyan2021characterizing} for this problem. In Figure \ref{fig:failure_mode} (B), $2,500$ collocation points are used, whereas, in the rest of the paper, $10,000$ are used for every nondimensionalized length of one in the temporal domain. When a larger number of points is used, we find we can consistently solve this problem with a standard PINN. Therefore, we classify this training challenge \cmnt{failure mode}by its apparent failure to propagate any information as the initial condition features abruptly stop, indicating the point density is too small. Overcoming this \cmnt{failure mode}challenge through increased and adaptive sampling is investigated in more detail in \cite{daw2022rethinking}. This allows for a constant solution to prevail in the rest of the domain. \\

\textbf{Incorrect Propagation:} Incorrect propagation \cmnt{failure mode}is reproducible by trying to solve the Allen-Cahn problem with a PINN, regardless of standard model tuning, as seen in Figure \ref{fig:failure_mode} (C). This challenge \cmnt{failures}arises when strong enforcement of causality is needed, such as in chaotic problems shown in \cite{wang2022respecting}, and by not enforcing it, the PINN converges to an incorrect solution. It is distinct from the zero-solution challenge \cmnt{failure}since a solution is arrived at quickly, but not the correct one.

\rmk{Interestingly, the training challenge \cmnt{failure mode}for long-time solution of the KdV problem is incorrect propagation instead of the zero-solution, such as in long-time convection. This result is described in \ref{ssec:kdv_long-appendix}.}

\subsubsection{Temporal Decomposition Challenges\cmnt{Failures}}
\label{ssec:temporal_failure}
\begin{figure}[H]
\centering
\includegraphics[width=1\linewidth]{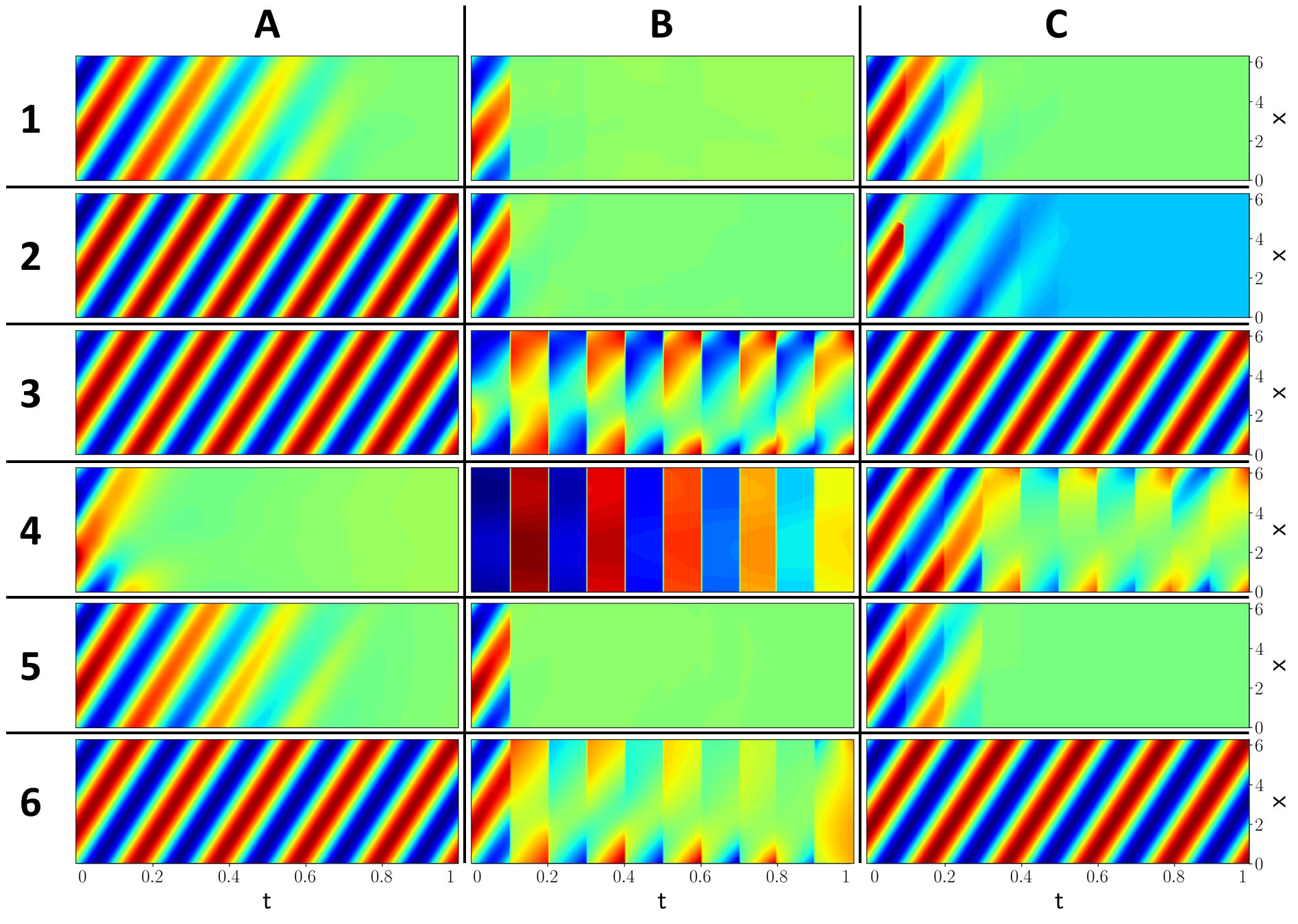}
  \caption{\small Convection problem on $T = [0, 1]$. (A) PINN. (B) XPINN. (C) XPINN with only the solution continuity interface \cite{li2022meta}. (1) 500 Adam + 2,500 L-BFGS. (2) 500 Adam + 10,000 L-BFGS. (3) 500 Adam + 2,500 L-BFGS + Dirichlet BC. (4) 500 Adam + Dirichlet BC. (5) 500 Adam + 2,500 L-BFGS + Weak BC. (6) 500 Adam + 2,500 L-BFGS + Dirichlet \& Weak BC.}
  \label{fig:failure_decomp}
\end{figure}

Let us now consider the convection problem with $T_{end} = 1$. In Figure \ref{fig:failure_decomp}, this PDE problem is run with a PINN (A), an XPINN (B), and an XPINN (more accurately, a multi-domain PINN \cite{li2022meta}) using only solution continuity conditions at the interfaces (C). All models contain the same point sets, loss term weights, etc., with the addition of interface sets in the decomposition models. Unless stated otherwise, exact periodic boundary enforcement is used with $M = 1$, as described in \ref{sec:fourier_appendix}, where $M$ is the order of the Fourier feature encoding.

In Figure \ref{fig:failure_decomp} (A.1), due to periodic boundary conditions, the solution propagates from the initial condition, whereas the rest of the domain converges to the zero-solution because it must satisfy the PDE residual but has no unique information. The collocation points inside the domain where information has not yet propagated provide no benefit despite taking computational time to compute the PDE residual, which can be crippling if the problem has high sampling density, is high dimensional, or is a long-time problem since the number of point-wise predictions and gradients is ever increasing. In the case of domain decomposition approaches like XPINNs and cPINNs, where all networks are trained at once, this can, in fact, cause training challenges \cmnt{failure modes}where there were none with a standard PINN, even though parallelization can help alleviate the training cost. To highlight this, in (A.2), the PINN is run for more L-BFGS iterations and converges appropriately.

In Figure \ref{fig:failure_decomp} (B.1) \& (C.1), the solution struggles to propagate information through the first interface (in this case, at dt = $0.1$). Since all networks are trained in concurrently from the start, the ones at later times become stuck in the local minima of the trivial zero-solution. This problem is intuitive to understand and is the same issue discussed in Section \ref{ssec:failure_mode_types} with respect to the long-time convection problem for a PINN. However, the issue is exacerbated here since, later in time, networks do not have direct access to the initial condition information; only through the interfaces, once the information has reached them, is a unique solution defined.
In (B.2) and (C.2), little has changed with the addition of more training iterations. The models will not overcome this challenge \cmnt{failure}with more training. Causality enforcement must be introduced to alleviate this issue. 

A standard XPINN (B), which has a residual continuity term, further intensifies training issues because the interface also has a zero-solution challenge. We claim that for temporal decomposition, $\mathcal{C}^p$ continuity should be used instead of the standard XPINN continuity conditions, which perform worse in all scenarios of our study. This effect worsens when using periodic conditions because it allows for the zero-solution more readily. Furthermore, it is the boundary condition most papers use that focuses on PINN ``failure mode'' problems \cite{krishnapriyan2021characterizing, wang2022respecting} despite not identifying it as a contributing factor. In (C.3), applying Dirichlet boundary conditions to the domain decomposed model with solution continuity allows for the correct solution to be obtained, whereas in (B.3), the XPINN interface conditions still cause propagation issues\cmnt{the failure mode}. To a lesser extent, it is also the case for PINNs that Dirichlet instead of periodic boundary conditions are easier to train, as seen in (A.3) which converges while (A.1) has not, despite equivalent training iterations. 

Finally, in (5) and (6), setups are repeated using weakly imposed, instead of exact (by way of Fourier feature encoding), periodic boundary conditions. Weakly imposed boundary conditions result in the same set of correct and incorrect solutions as before. Previous work implies that exact enforcement of periodic conditions may alleviate training issues, but we find that regardless of the enforcement, the problems can persist. Only different boundary conditions, such as Dirichlet, change the result. 

For these reasons, time-marching, with the same amount and density of collocation points, helps alleviate the trivial zero-solution trivial \cmnt{failure}for temporal decomposition and can be described under the lens of information propagation. Time-marching, in effect, removes the collocation points later in time from optimization, not allowing the model to train itself into a trivial solution later in time, even though multiple subnetworks are similarly used in XPINNs. The resistance to propagate information is an optimization and uniqueness issue, as the null-space is an acceptable solution to the optimization problem. Despite the PDE being violated in between the true and zero-solution, it does not train out of the local minima. In the case of domain decomposition, the interface is an ideal location to violate the PDE and stop information from propagating. Whereas for a PINN on the long-time convection problem, this violation happens over a large time span as the feature gradually weakens. 

\rmk{Information propagation is not fully understood and depends on multiple PINN aspects, such as the optimizer, sampling method, etc., which are not all studied here. For example, in the 3D Euler equation, characteristic information is complicated, making methods such as LPINN \cite{mojgani2022lagrangian} and CINN \cite{braga2022characteristics}, difficult.}

\section{Unified Causality-Enforcing Framework}
\label{sec:methods}

To address these decomposition challenges \cmnt{failure modes}and unify previous causal strategies, we propose two new methods to cover all aspects of causality enforcement shown in Table \ref{table:causality-class2}. Combined, these two methods impose soft and hard constraints on both the time-slab and sampling scale. We also introduce ways to improve temporal decomposition, such as transfer learning. 

\begin{center}
\captionof{table}{\small A classification of PINN causality enforcement methods with our proposed stacked-decomposition and window-sweeping methods.} 
\scalebox{0.8}{
\begin{tabular}{| c | c | c | c | c |}
\cline{2-5}
\multicolumn{1}{c|}{} & \textbf{Soft Causality} & \textbf{Hard Causality} & \textbf{Soft + Hard Causality} & \textbf{non-Causal}\\
\hline
\textbf{Time-slab scale} & Adaptive time-sampling \cite{wight2020solving} & Time-marching \cite{wight2020solving, krishnapriyan2021characterizing, bihlo2022physics} & \textit{Stacked-decomposition} & XPINN \cite{JagtapK} \\
& & bc-PINN \cite{MATTEY2022114474} & & \\
\hline
\textbf{Sampling scale} & Causal weights\cite{wang2022respecting} & - &  \textit{Window-sweeping} & - \\
\hline
\end{tabular}}
\label{table:causality-class2}
\end{center}

\subsection{Stacked-decomposition}

\begin{figure}[H]
\centering
\includegraphics[width=0.85\linewidth]{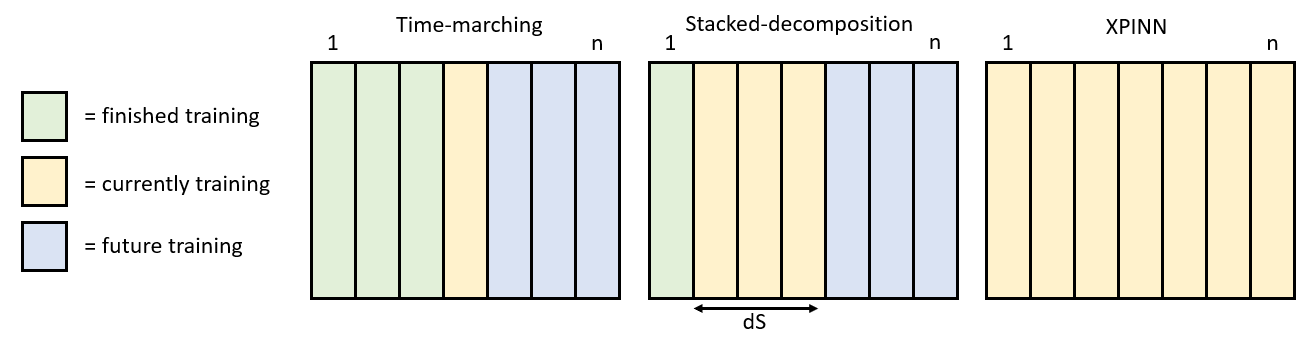}
  \caption{\small Illustration of the proposed stacked-decomposition method compared with the existing time-marching and XPINN methods. }
  \label{fig:stacked-xpinn}
\end{figure}

As seen in Figure \ref{fig:stacked-xpinn}, stacked-decomposition is parameterized by $n$ and $dS$. The length that a subdomain spans in time is then inferred from the total time domain for each problem and the number of partitions $n$. For $dS = 1$, stacked-decomposition is equivalent to time-marching. For $dS = n$ with XPINN interface conditions and all domains active at the start of training, stacked-decomposition is equivalent to the traditional XPINN approach. An additional term we define is causal $dS$: which describes if the amount of networks $dS$ represents should all be trainable at the start or if a warm-up procedure is used (starting at one and increasing to $dS$). When used with $dS = n$, we refer to this model as a ``causal XPINN''. In this configuration, later time-slabs are added as the prior slab reaches convergence, and the entire set of subnetworks continues to train. A causal XPINN arrives at the standard XPINN configuration once all subnetworks have been added. However, because of the warm-up procedure, it avoids the training challenge \cmnt{failure mode}described in Section \ref{ssec:motivation}. This is because future networks do not train to the zero-solution and are only added once the information in the previous slab has propagated to the final time in the subdomain. A main benefit of XPINNs is that they can be parallelized and, therefore, handle large-scale problems. In this regard, as subnetworks are added to causal XPINNs, they can be parallelized, introducing no limitation or cost. This contrasts time-marching, in which all prior networks must conclude training and run in sequence. Therefore, stacked-decomposition can describe an ideal middle ground in which we benefit from the causality of time-marching to avoid possible training difficulties \cmnt{failure modes}and the parallel training of XPINNs. The method also describes a new set of models when $1 < dS < n$, which may be useful for large-scale problems with time-history effects where training the full domain at once is expensive, but the information in prior domains is still useful. In the future, adaptive methods for determining $n$ a priori or during training will be considered since time scale correlation or local complexity may change with time. 

\subsubsection{Interface Conditions}
Attempting to bridge the gap between temporal decomposition strategies, we must explain the differences in interface conditions in the loss term. Time-marching schemes use the final time prediction of the previous time-slab as the initial condition of the next time-slab. For first-order time problems, this condition is simply the solution continuity given by
\begin{align}
&\mathcal{L}_{i}(\boldsymbol{\theta}^{-}, \boldsymbol{\theta}^{+}) = \frac{1}{N_i} \sum_{i=1}^{N_i} \vert u_{\boldsymbol{\theta}^{-}}(x_i, t) - u_{\boldsymbol{\theta}^{+}}(x_i, t) \vert^2.
\label{eq:interface-ic} 
\end{align}
We generalize this and refer to it as the $\mathcal{C}^p$ continuity where $p$ is the order in time minus one. For problems considered in this paper, it will be $\mathcal{C}^0$ and, as such, equivalent to the solution continuity. Traditional XPINNs use interface conditions of discontinuous solution continuity and residual continuity given by the following loss terms:
\begin{align}
\begin{split} 
&\mathcal{L}_{i_{avg}}(\boldsymbol{\theta}^{-}, \boldsymbol{\theta}^{+}) = \frac{1}{N_i} \left( \sum_{i=1}^{N_i} \left( \vert u_{avg}(x_i, t) - u_{\boldsymbol{\theta}^{+}}(x_i, t) \vert^2 + \vert u_{avg}(x_i, t) - u_{\boldsymbol{\theta}^{-}}(x_i, t) \vert^2 \right) \right) \\
& \quad \quad \equiv \mathcal{L}_{i_{avg}}(\boldsymbol{\theta}^{-}, \boldsymbol{\theta}^{+}) = \frac{1}{2N_i} \sum_{i=1}^{N_i} \vert u_{\boldsymbol{\theta}^{-}}(x_i, t) - u_{\boldsymbol{\theta}^{+}}(x_i, t) \vert^2 \leftarrow u_{avg} = \frac{u_{\boldsymbol{\theta}^{-}} + u_{\boldsymbol{\theta}^{+}}}{2} \label{eq:interface-avg} 
\end{split} 
\\
&\mathcal{L}_{i_{\mathcal{R}}}(\boldsymbol{\theta}^{-}, \boldsymbol{\theta}^{+}) = \frac{1}{N_i} \sum_{i=1}^{N_i} \vert \mathcal{R} \left( u_{\boldsymbol{\theta}^{-}}(x_i, t) \right) - \mathcal{R} \left( u_{\boldsymbol{\theta}^{+}}(x_i, t) \right) \vert^2.
\label{eq:interface-res}
\end{align}
However, the discontinuous continuity reduces to the continuous continuity with a constant, and given that tuning loss terms and weights have been extensively studied and are part of the XPINN framework \cite{wang2022and, JagtapK}, we will make no distinction between these two terms as loss term weighting will override the factor of one half difference. Finally, since we are decomposing in time, there is no complex geometry with which we must compute the normal, such as in cPINNs \cite{JAGTAP2020113028}. Therefore, residual continuity is not necessary in time since we can use the solution continuity, which is equivalent to the initial conditions for a new domain and makes the problem well-posed. Gradient-based interface terms may also become prohibitively expensive as the number of concurrently trained subdomains increases. However, it may be helpful in training to include multiple interface terms as studied in \cite{li2022meta}, so it is left up to the user and the problem to define which terms to include, such as residual continuity, so long as they are well-posed. 
\rmk{Straight lines for time-slabs are used for convenience since it is common for time-marching schemes. However, if an irregular shape is used, the same $\mathcal{C}^p$ continuity can be used and is still well-posed without any modification.}

\subsubsection{Transfer learning}
Transfer learning fits naturally into our framework when multiple networks are stacked sequentially in time. A variation of this application was used in \cite{bihlo2022physics} for time-marching. However, it was only briefly touched upon and not thoroughly studied as we do here. We further state there is no need to retrain the network from scratch when a network that already obeys the initial or interface condition is known. In terms of stacked-decomposition, it is easy to see that regardless of $dS = 1$, in which case there are initial conditions, or $dS > 1$, in which case there are interface conditions, initializing the following network with the prior network will result in this term being exactly zero when added. This aspect goes beyond simply having a good starting point for optimization since we are transferring to a new domain that shares predictions with the model being transferred. Residual loss terms beyond the starting subdomain time will not be zero, as this region will be an extrapolation of the prior subdomain. However, it will be closer to convergence than randomizing the weights. 

\begin{figure}[H]
\centering
\includegraphics[scale = 0.3]{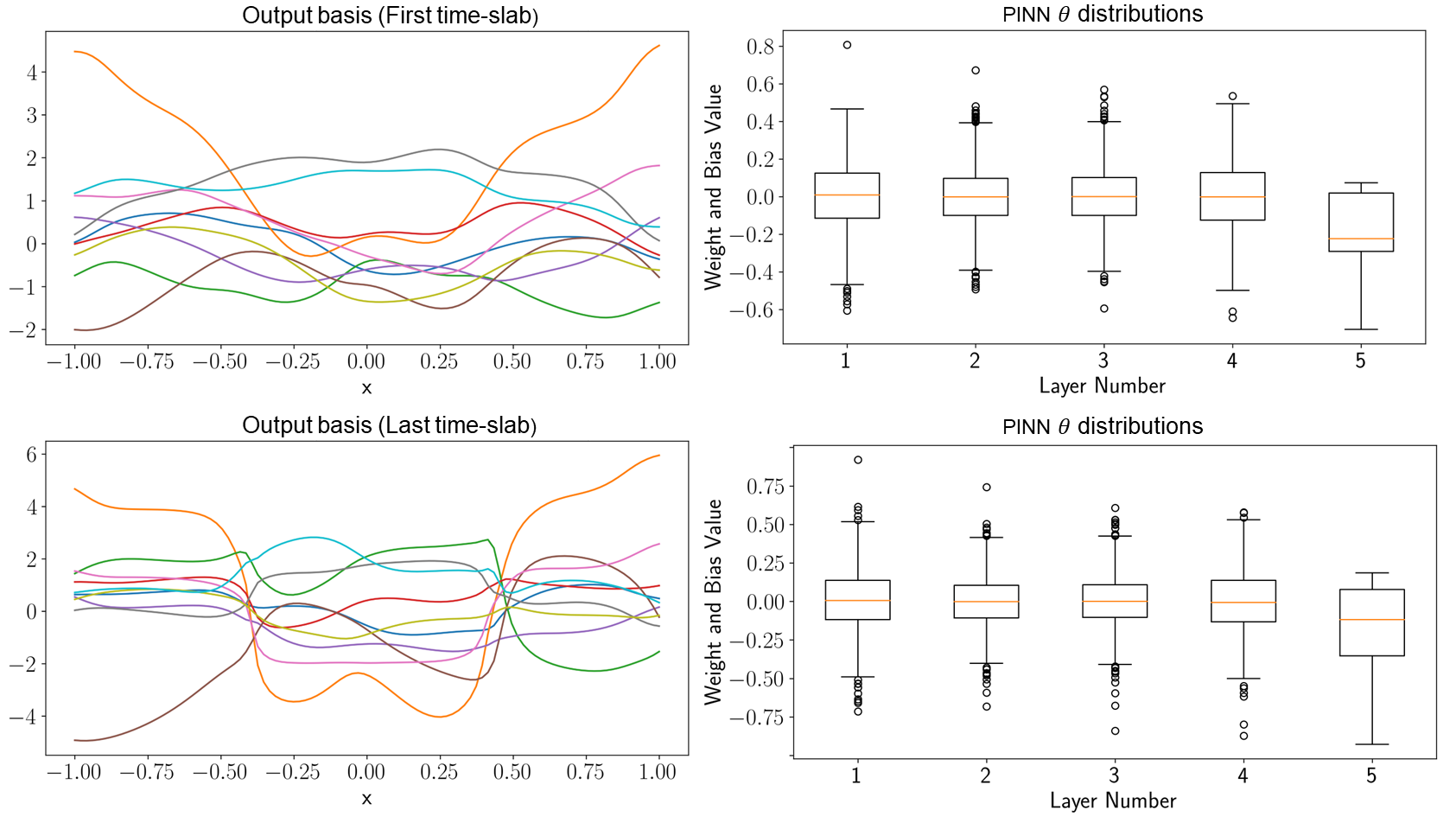}
  \caption{\small Spatial basis at the center of the time-slab given by the final layer of the PINN with time-marching on the Allen-Cahn problem in \cite{raissi2019physics}. The basis changes considerably between the first and last time-slabs, indicating true transfer learning would not work as the scales change in time for this problem. The distribution of learnable parameters is also shown not to change significantly despite the change in basis.}
  \label{fig:transfer-basis}
\end{figure}

In this framework, we allow the flexibility of transferring any combination of layers and holding constant any combination of transferred layers. More precisely, we define the terms ``transfer learning'' and ``fine tuning'' to aid in this explanation. Traditionally, transfer learning refers not only to initializing the learnable parameters of one network with another but also to holding some number of the layers constant, which reduces the per-iteration cost. On the other hand, we will refer to fine tuning as the initialization of learnable parameters while still allowing the full network to be trainable. We claim this is an important distinction given this application because scales and solution dynamics may change over time, meaning that holding some layers constant may inhibit the expressibility of the network and its ability to accurately fit the true solution. Let us take the final linear combination of the network as basis functions and consider the nonlinear Allen-Cahn problem in \cite{raissi2019physics}. For the time-marching model, it can be seen that the basis sharpens from the first to the final subdomain in Figure \ref{fig:transfer-basis}. 

The overall network parameter distribution for each layering stays close to constant despite the drastic change in output basis, meaning this alone is not a good indicator of what is being learned. While fine tuning can still improve training in this case, transfer learning would inhibit it as we need earlier layers in the network to change so that the final basis can more accurately fit the smaller scales that form as time goes on in this problem.   

\subsection{Window-sweeping collocation points}

\begin{figure}[H]
\centering
\includegraphics[scale=0.4]{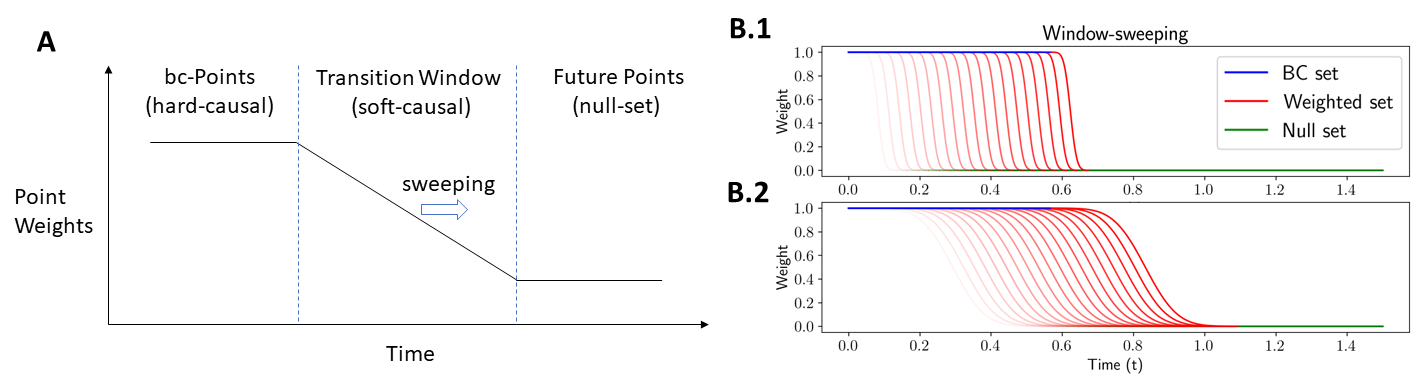}
  \caption{\small (A) Illustration of window-sweeping method and its corresponding collocation point subsets. (B) Window-sweep propagation over training time. (B.1) Error function kernel with a steep transition. (B.2) Error function kernel with a smooth transition.}
  \label{fig:window-sweep}
\end{figure}

As seen in Figure \ref{fig:window-sweep}, a soft causality window is moved through time, which acts as a weight mask on the collocation points. Unlike stacked decomposition, this method is defined by a set of point weights moving forward in time in a single PINN. We find this view can describe many previous and new methods.

Inspired by causal weighting in \cite{wang2022respecting}, this transition can be defined in many ways, which we will colloquially call the kernel. One option is to define it using the causal weighting scheme but to add upper and lower bound cutoffs to move those points into the prior time set of backward-compatibility points and the future set of points that have not yet been included in the training. The backward-compatibility set acts as a hard causality constraint in addition to the computational benefit of not requiring the expensive PDE residual. Causal weights have shown great performance on difficult PDE problems; however, they set future collocation point weights to zero until prior residuals have been satisfied, wasting time predicting and computing gradients for points that contribute negligibly to the overall loss landscape and, therefore, optimization. The inclusion of the null-set bound removes this inefficiency until the points are useful. In the user algorithm, the addition or absence of these sets is variable so that the causal weights method can be recovered. Since causal weights are explicitly based on prior residuals, this cutoff on the upper bound would be known without having to perform any operations on future points and, therefore, incur no additional cost. Other kernels considered in this paper are shown in Table \ref{tb:window-kernel}. Depending on the problem and hardware capacity, larger or smaller weighted domains can be considered, as shown in Figure \ref{fig:window-sweep} (B) with the error function kernel. Using the uniform weight kernel, bc-PINNs can be recovered when width is set to dt. In the case of spatial information propagation in multiple directions, such as reflecting boundary conditions, no modification to this method is necessary as it is purely time-dependent. However, in future work, a method similar to window-sweeping that acts spatially could significantly improve training and complement the methods presented herein. Other future work will consider modifying this method to solve second-order time problems with initial and final conditions on $u$ that require information to propagate in both directions. 

\begin{center}
\captionof{table}{\small Window-sweeping kernel hyperparameters. The dt tolerance, similar to the tolerance in stacked-decomposition, is a bound on the change in loss required for the point-set bounds to move in time by the defined dt. This is analogous to wave speed but for information propagation as a function of PINN training. ``Offset'' in the ``Analytical Form" column is a complex term representing the propagation of the window throughout training, which is dependent on terms such as dt tolerance (see B.1 \& B.2 in Figure \ref{fig:window-sweep} for visualization).}

\scalebox{0.85}{
\begin{tabular}{| c | c | c |}
\hline
Kernel & Hyperparameters & Analytical Form of Weighting\\
\hline
Uniform & [width, dt, dt tolerance, scale] & $\begin{cases}
    \text{scale} & \text{if } t \leq \text{width} + \text{offset}\\
    0   & \text{if } t > \text{width} + \text{offset}\\
\end{cases}$\\
Linear & [width, dt, dt tolerance, scale] & $\begin{cases}
    \text{scale} & \text{if } t < \text{offset}\\
    \text{scale} \cdot \frac{-t}{\text{width}} & \text{if } \text{offset} \leq t \leq \text{width} + \text{offset}\\
    0   & \text{if } t > \text{offset}\\
\end{cases}$\\
Error Function & [steepness, dt, dt tolerance, scale, cutoff tolerance] & $\text{scale} \cdot erf(\text{steepness} \cdot (-t + \text{offset}))$ \\
Causal Weights & [$\epsilon$, cutoff tolerance] & $exp(-\epsilon \sum^{n}_{i} \mathcal{L}_r (t_i, \boldsymbol{\theta}))$\\
\hline
\end{tabular}}
\label{tb:window-kernel}
\end{center}

\subsection{User Settings}
The conglomeration of these methods form a procedural framework by which we attempt to capture as many temporal PINN training techniques as possible as a subset of the options. Additionally, the algorithm allows for a full range of variants, combinations, and improvements. The procedure consists of the following high-level steps:
\begin{enumerate}[\indent {}]
  \item {\bf Step 1:} Choose stacked-decomposition parameters \textbf{[$\mathbf{n}$, $\mathbf{dS}$, \textbf{causal} $\mathbf{dS}$, \textbf{tolerance}]}; 
  \item {\bf Step 1.1:} Choose interface conditions \textbf{[residual continuity, $\mathbf{\mathcal{C}^p}$ continuity, other]};
  \item {\bf Step 1.2:} Choose transfer learning parameters \textbf{[number of layers, trainability of layers]}; 
  \item {\bf Step 2:} Choose window-sweeping parameters \textbf{[bc-set, null-set, weighting kernel, kernel hyperparameters]}
\end{enumerate}

To highlight versatility, we will define the existing models listed in Section \ref{ssec:related-work} in terms of procedural choices. A subtle but large improvement is in the addition of a tolerance to stacked-decomposition, which the user sets to define the change in loss before a new subdomain is added. This minimizes the cost of unnecessary training time used in the original papers for time-marching, bc-PINNs, etc., that evaluated a fixed number of iterations before moving to the next time-slab. Using a tolerance also reduces hyperparameter tuning, as an underestimate of iterations may lead to an incorrect solution\cmnt{training failure} and an overestimate is expensive. 

\rmk{Other methods such as adaptive weighting and sampling techniques (self-adaptive weights, RAR, Evo, and self-supervision adaptive sampling \cite{mcclenny2020self, lu2021deepxde, daw2022rethinking, subramanian2022adaptive}), or reformulating the network architecture to obey characteristics (LPINN, CINN \cite{mojgani2022lagrangian, braga2022characteristics}) can be used along with this framework, but do not fall into our unification of like methods.}

\begin{center}
\captionof{table}{\small Existing PINN methods and their corresponding settings under the proposed framework.} 
\scalebox{0.85}{
\begin{tabular}{| c | c | c | c | c |}
\hline
Existing Method & Step 1. & Step 1.1 & Step 1.2 & Step 2. \\
\hline
PINN & [1, 1, off, Any] & [None] & [None] & [None] \\
Adaptive time-sampling & [1, 1, off, Any] & [None] & [None] & [off, on, uniform, width = dt, scale = 1]\\
Time-Marching & [n, 1, off, Any] & [$\mathcal{C}^p$] & [None] & [None] \\
bc-PINN & [1, 1, off, Any] & [None] & [None] & [on, on, uniform, width = dt, scale = 1]\\
Causal weights & [1, 1, off, Any] & [None] & [None] & [off, off, causal weights, $\epsilon$]\\
XPINN & [n, n, off, Any] & [Residual, $u_{avg}$] & [None] & [None]\\
\hline
\end{tabular}}
\label{table:algo-subsets}
\end{center}

Additionally, a code package is included with this paper, which allows for easy configuration of options for new and existing problems using PyTorch for first-order in time PDEs. \footnote{The code and data accompanying this manuscript will be made publicly available at \url{https://github.com/mpenwarden/dtPINN} after publication.}

\section{Numerical Experiments}
\label{sec:results}
In this section, we demonstrate the efficacy of our proposed framework on various forward PDE problems. With these results, we seek to highlight the flexibility and variability of our framework in easy-to-define models with simple user settings. We do not advocate for one method over another in terms of accuracy or runtime but rather provide a thorough comparison of a subset of all possible choices. Ground truth solutions are generated using the Chebfun package \cite{Driscoll2014} with a spectral Fourier discretization with 512 modes and a fourth-order stiff time-stepping scheme (ETDRK4) \cite{COX2002430} with time-step size $10^{-5}$. The training set is composed of $10,000$ residual collocation points ($N_r$) using Latin hypercube sampling (LHS) and $200$ uniformly spaced boundary points ($N_b$) for every nondimensionalized length of one in the temporal domain. Each neural network is comprised of $50$ neurons and $4$ hidden layers. The collocation set is chosen using Latin-hypercube sampling. The initial condition and each interface consist of $200$ uniformly spaced points ($N_{ic}$ \& $N_i$). All models use Fourier feature encoding, described in \ref{sec:fourier_appendix} unless weak boundary conditions are stated. If Fourier feature encoding is not used, the spatiotemporal input is normalized between $[-1, 1]$. Casual $dS$ is used for all stacked-decomposition models unless otherwise stated. The total loss for any given model can be written as
\begin{align}
MSE = \lambda_r MSE_r + \lambda_{BC} MSE_{BC} + \lambda_{IC} MSE_{IC} + \lambda_{bc} MSE_{bc} + \lambda_{i} MSE_{i} 
\end{align}
where $MSE_{\#}$ is $0$ if unused, and $\lambda_r = 1$, $\lambda_{BC} = \lambda_{IC} = \lambda_{bc} = \lambda_{i} = 100$ unless stated otherwise. These experiments were run on an Intel Core i7-5930K processor with Windows 10 OS. The test performance is reported in relative $L_2$ error given by
\begin{align}
	\frac{||u-u_{\theta}||_2}{||u||_2}
\end{align}
as well as wall-clock training time. 

\rmk{For both stacked-decomposition and window-sweeping methods, loss tolerances can be decreased to potentially gain accuracy at the cost of additional training time. The parameter choices made provide a reasonable trade-off. As with all machine learning methods, the choice of tunable hyperparameters will depend on the intended use, and the results reported cannot be completely exhaustive of all training possibilities. Our goal is to make overarching insights, not tell the user the correct settings in each scenario.}

\subsection{Convection equation}
\begin{figure}[H]
\centering
\includegraphics[width=1\linewidth]{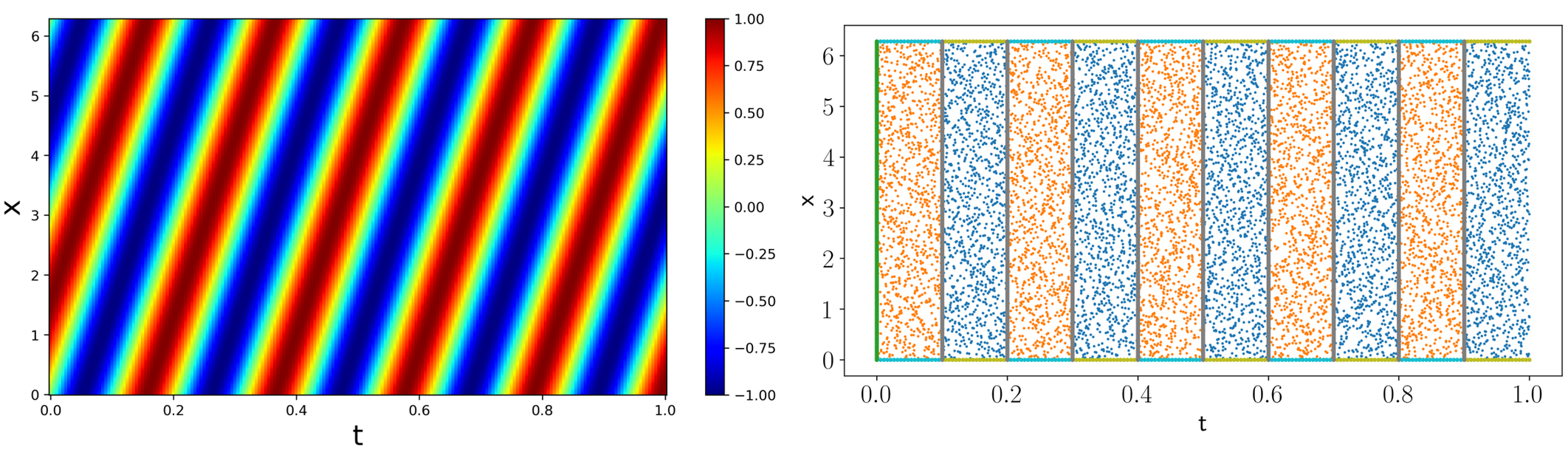}
  \caption{\small (Left) Exact solution. (Right) Plot of 10 subdomains delineating the individual initial condition, boundary condition, interface, and subdomain collocation point sets.}
  \label{fig:convec_exact}
\end{figure}

Let us consider the following convection problem
\begin{linenomath}\begin{align}
& \frac{\partial u}{\partial t} + 30\frac{\partial u}{\partial x} = 0, \; (t,x)\in [0,1] \times [0,2 \pi]
\label{eq:convection}
\end{align}\end{linenomath}
subject to periodic boundary conditions and an initial condition $u(0,x) = sin(x)$. The exact solution and point sets are shown in Figure \ref{fig:convec_exact}.

\begin{table}[H]
	\centering
	\captionsetup{width=1\linewidth}
    \caption{\small Table of $L_2$ relative error and training time for different stacked-decomposition settings. M = 1 unless weak boundary conditions are used. Note stacked-decomposition is abbreviated s-d, interface condition as ic, residual continuity as rc, fine tuning as FT, and transfer learning as TL.}
	\begin{tabular}[c]{l | c | c}
	\toprule
        Model settings & Relative $L_2$ Error & Training time (sec) \\
        \hline
        PINN & $8.28 \times 10^{-3}$ & 1,020 \\
        PINN + weak BC & $2.94 \times 10^{-2}$ & 780 \\
        s-d PINN (n = 10, dS = 1, ic = $C^p$)  & $1.23 \times 10^{-2}$ & 1,141 \\
        s-d PINN (n = 10, dS = 3, ic = $C^p$)  & $4.47 \times 10^{-3}$ & 4,240 \\
        s-d PINN (n = 10, dS = 1, ic = $C^p$)  + weak BC & $7.69 \times 10^{-2}$ & 547 \\
        s-d PINN (n = 10, dS = n, ic = $u_{avg}$ + rc) + FT & $3.90 \times 10^{-2}$ & 21,443 \\
        s-d PINN (n = 10, dS = 1, ic = $C^p$) + FT  & $7.43 \times 10^{-3}$ & 703 \\
        s-d PINN (n = 10, dS = 3, ic = $C^p$) + FT & $5.13 \times 10^{-3}$ & 2,261 \\
        s-d PINN (n = 10, dS = n, ic = $C^p$) + FT  & $4.11 \times 10^{-3}$ & 5,066 \\
                s-d PINN (n = 10, dS = 1, ic = $C^p$) + FT + weak BC & $3.44 \times 10^{-2}$ & 420 \\
        s-d PINN (n = 10, dS = 1, ic = $C^p$) + TL  & $1.96 \times 10^{-2}$ & 1,342 \\
        s-d PINN (n = 10, dS = 1, ic = $C^p$) + TL + weak BC & $1.62 \times 10^{-2}$ & 490 \\
	\bottomrule
	\end{tabular}
    \label{tb:convec}
\end{table}

In Table \ref{tb:convec}, many variations of stacked-decomposition are run for the convection problem. First, a standard PINN is able to solve the problem with relatively good accuracy and cost. We also observe that, unlike all results for the standard XPINN in Section \ref{fig:failure_decomp}, the causal XPINN with fine tuning (Table line 6) can converge to the correct solution, albeit with great computational cost. Therefore, we have demonstrated that even with the most unfavorable conditions, such as periodic boundaries and XPINN interfaces, causal enforcement, and transfer learning are able to overcome the zero-solution issue\cmnt{failure mode}.

Another result is that, $dS$ = 1 to $dS$ = $n$ acts as a spectrum of trade-off between accuracy and cost. Looking at the results with fine tuning applied, $dS$ = 1, which is equivalent to time-marching, converged the fastest since only one network is training at once, lowering the cost. As $dS$ increases to three and then $n$, the cost increases, but the accuracy improves since training networks concurrently allows them to better resolve the solution and any discrepancies at the interfaces. Distributed parallel training \cite{shukla2021parallel} can reduce this additional cost while retaining improved accuracy. 

We observe that weak boundary condition enforcement takes less time to reach convergence and is significantly less accurate. We also observe that true transfer learning is not appropriate for temporal decomposition, but fine tuning is. This issue is described in more detail in \ref{sec:ft_vs_tl_appendix}. In summary, stacked-decomposition, particularly with $dS$ = 1 and fine tuning, can outperform the standard PINN in accuracy and cost. This is significant as even for problems in which the unmodified PINN does not fail, the framework improves scalability in PINNs and yields improvement even on a short-time problem with relatively small amounts of points and training.

\begin{figure}[H]
\centering
\includegraphics[width=1\linewidth]{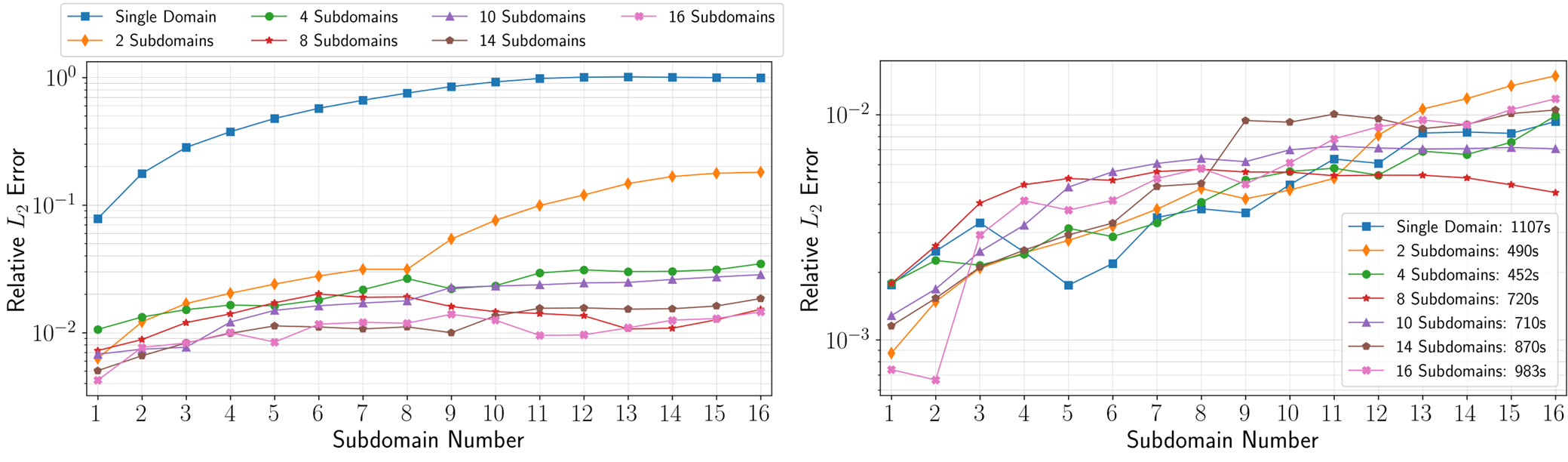}
  \caption{\small Relative $L_2$ error in 16 subdomains for various numbers of decomposition partitions. (Left) Adam optimizer only until convergence. (Right) 500 Adam warm-up iterations, then L-BFGS optimization until convergence.}
  \label{fig:convec_convergence}
\end{figure}
To investigate the effect of increasing the number of subdomains in causal, temporal decomposition, we systematically compare the relative $L_2$ error over subdomain sets for various settings. Starting with the single domain (PINN), we decompose the domain into n = 2, 4, 8, 10, 14, and 16 subdomains uniformly in time and report the relative $L_2$ error for each in 16 uniform subdomains. The temporal decomposition strategy is s-d PINN (n = \#, dS = 1, ic = $C^p$) + FT. Distinct from other experiments performed, we also consider optimizer choice in this study to provide insight into a main point of contention in PINNs training, Adam vs. L-BFGS. In Figure \ref{fig:convec_convergence} (Left), it is clear that for a single-domain PINN with only Adam optimization, the loss function gets stuck in a suboptimal local minima. As we introduce more subdomains, the relative $L_2$ error decreases. Eventually, the relative error converges, i.e., there is no improvement in predictive accuracy even after further decomposing the subdomain. Therefore, we observe that causal, temporal decomposition can overcome training challenges \cmnt{failures}due to poor optimizer choice, as well as previously discussed ones \cmnt{failures}in Section \ref{ssec:motivation}.

Figure \ref{fig:convec_convergence} (Right) uses a warm-up of 500 Adam iterations before switching to L-BFGS. This warm-up is known to reduce the failure of L-BFGS in the early stage of training. In contrast to Adam only optimization, the error is relatively constant throughout the number of subdomains. We report training times in this case because all methods converge. Training times are not reported for Adam only training since it is misleading to analyze when some cases fail and some do not. We can see that even when training challenges \cmnt{failure mode}are not present, causal, temporal decomposition can improve training time and, therefore, the scalability of PINNs in larger and more expensive problems. However, there appears to be an ideal subdomain number, which will be problem specific, and going beyond what is necessary increases run time with no benefit here. This is likely due to the interplay between the cost of refined learning of the network when the loss changes slowly, which must happen in all subnetworks, versus the benefit of convergence speed for smaller domains. 



\subsection{Allen-Cahn equation}
\begin{figure}[H]
\centering
\includegraphics[width=1\linewidth]{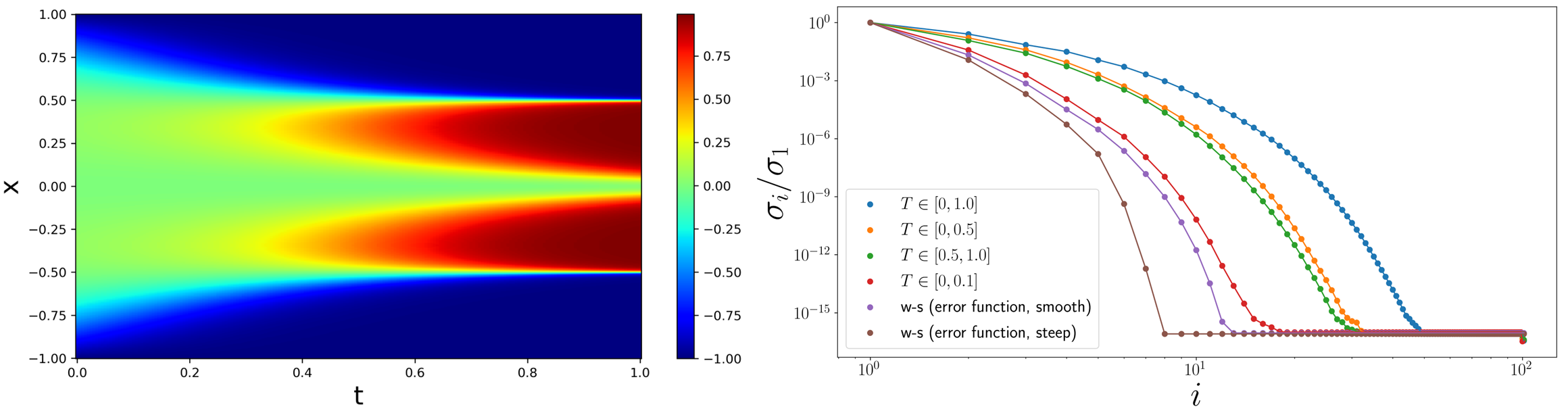}
  \caption{\small (Left) Exact solution. (Right) Study of (normalized) 
 singular value spectra of temporal snapshots (formalized by \cite{mojgani2022lagrangian}) for the Allen-Cahn problem.}
  \label{fig:exact_allen_cahn}
\end{figure}

Let us consider the following Allen-Cahn problem
\begin{linenomath}\begin{align}
& \frac{\partial u}{\partial t} - 0.0001\frac{\partial^2 u}{\partial x^2} + 5u\left(u^2-1\right) = 0, \; (t,x)\in [0,1] \times [-1,1]
\label{eq:allen-cahn}
\end{align}\end{linenomath}
subject to periodic boundary conditions and an initial condition $u(0,x) = x^2 cos(\pi x)$. In Figure \ref{fig:exact_allen_cahn} shows the exact solution and (normalized) singular value spectra of temporal snapshots for different data sets representative of decomposition and point weighting schemes. The lens of Kolomogrov n-widths, approximated by the rate of decay of these singular values, is proposed as an \textit{a priori} PINNs convergence estimate in \cite{mojgani2022lagrangian}. We use this lens to view the Allen-Cahn problem with different time-slab sizes in addition to the window-sweeping weighting scheme with the error function kernel. As described in \cite{mojgani2022lagrangian}, a faster decay rate of the singular values of a set of snapshots should correlate to an increase in the rate of training convergence. We observe that smaller time-slabs have faster decay, which empirically aligns with faster training, potentially leading to reduced training times. The smooth error function kernel corresponds with zero-valued weights past $t = 0.1$. Therefore, compared to the decay $t \in [0, 0.1]$, which has no weightings over this region, the window-sweeping method has a faster drop-off, indicating it is even easier to train. 


\begin{table}[H]
	\centering
	\captionsetup{width=1\linewidth}
    \caption{\small Table of $L_2$ relative error and training time for different window-sweeping settings. All methods use M = 10 unless otherwise stated. The loss tolerance used to propagate all methods is $10^{-7}$. Note that window-sweeping is abbreviated w-s. $^a$(bc-set = on, null-set = on) $^b$(bc-set = off, null-set = on).} \label{tb:allen-cahn}
	\begin{tabular}[c]{l | c | c}
	\toprule
        Model settings & Relative $L_2$ Error & Training time \\
        \hline
        PINN & $5.11 \times 10^{-1}$ & 3,421 \\
        s-d PINN (n = 10, dS = 1, ic = $C^p$) + FT & $2.77 \times 10^{-2}$ & 798 \\
        w-s PINN (kernel = uniform, width = dt = $0.1$)$^b$  & $6.57\times 10^{-2}$ & 875 \\
        w-s PINN (kernel = uniform, width = dt = $0.1$)$^a$: M = 1  & $2.25\times 10^{-2}$ & 448 \\
        w-s PINN (kernel = uniform, width = dt = $0.1$)$^a$  & $1.73\times 10^{-2}$ & 466 \\
        w-s PINN (kernel = uniform, width = $2$dt = $0.1$)$^b$  & $3.33\times 10^{-2}$ & 1,053 \\
        w-s PINN (kernel = uniform, width = $2$dt = $0.1$)$^a$  & $1.58\times 10^{-2}$ & 574 \\
        w-s PINN (kernel = linear, width = $4$dt = $0.1$)$^a$  & $3.45\times 10^{-2}$ & 994 \\
        w-s PINN (kernel = error function, steep, dt = $0.0125$)$^a$  & $3.62\times 10^{-2}$ & 534 \\
        w-s PINN (kernel = error function, smooth, dt = $0.0125$)$^a$  & $4.29\times 10^{-2}$ & 564 \\
	\bottomrule
	\end{tabular}
\end{table}

In Table \ref{tb:allen-cahn}, many variations of window-sweeping are run for the Allen-Cahn problem. Unlike the convection problem considered, an unmodified PINN does not sufficiently solve this. The third row setting recovers adaptive time-sampling, and the fifth row recovers bc-PINNs as described in Table \ref{table:algo-subsets}. First, we find that all methods are able to overcome the training challenge encountered by the unmodified PINN. We also find that by adding the backward compatibility set instead of continuously training on prior point sets vastly decreases the training time with no adverse effect on the accuracy. 

Uniform weights perform well compared to soft causality enforcement via weighting schemes used by methods such as causal weights \cite{wang2022respecting} and extended to the unified window-sweeping method by way of kernels linear, error function, and an equivalent causal weighting scheme. Under the loss tolerance setting of $10^{-7}$ used, the causal weights kernel reaches this tolerance without sufficient training. Due to the sensitivity of its tunable causality parameter ($\epsilon$), as noted in the original paper, we present self-contained results for this kernel in \ref{sec:causal_weights_appendix}. We also extend the method to non-grid sampling and reduce training time using the null-set segmentation of window-sweeping. We also find that for uniform weights, reducing the dt size such that new sets overlap with prior slightly improves accuracy with increased cost. The primary motivation for the model settings reported is to showcase how simple it is to modify the proposed framework to produce new models, not to conclude which method is the ``best'' since different settings may be ideal for different problems. We also note the improved scalability of this approach, particularly in the application of the change in loss tolerances to propagate the methods. As a comparison, in bc-PINNs \cite{MATTEY2022114474}, the authors use 50,000 Adam iterations per segment and then L-BFGS iterations until tolerance termination, leading to hundreds of thousands of iterations. We report almost identical relative $L_2$ errors and use a total of around 12,000 iterations. This modification, used in both stacked-decomposition and window-sweeping, allows us to achieve more accurate solutions than unmodified PINNs in less time. 

\begin{figure}[H]
\centering
\includegraphics[width=1\linewidth]{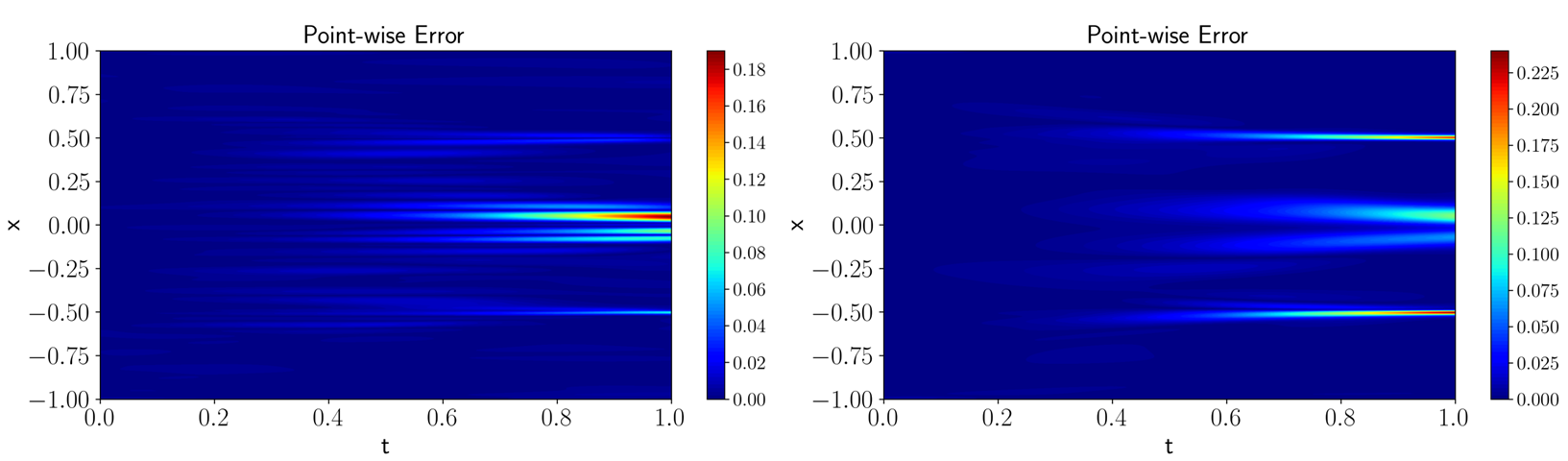}
  \caption{\small Point-wise error of w-s PINN (kernel = uniform, width = dt = $0.1$)$^c$ reported in Table \ref{tb:allen-cahn} (Left) M = 10 (Right) M = 1.}
  \label{fig:w-s_M1vsM10}
\end{figure}

To investigate the effect of Fourier feature encoding frequency, we run the window-sweeping model with equivalent settings to recover bc-PINNs using an encoding of M = 1 and M = 10 shown in Figure \ref{fig:w-s_M1vsM10}. This encoding is used in the paper introducing causal weights \cite{wang2022respecting} with M = 10. We find that a higher order encoding can better resolve sharper features, similar to adaptive activation functions \cite{jagtap2020adaptive}, the error manifests itself elsewhere at this fidelity of training. As seen on the left side of the figure, the discontinuities that begin to form at the end of time around $\pm 0.5$ have low point-wise error for $ M = 10$, although the error manifests itself in the relatively smooth $x = 0$ region. This is opposed to $M = 1$, which struggles at the discontinuities. For higher levels of training, the higher order encoding will help resolve smaller scales in the solution space. However, at these stopping tolerances, we report similar accuracies for both encoding choices. 

\subsection{Korteweg–de Vries equation}
\label{ssec:kdv}
\begin{figure}[H]
\centering
\includegraphics[width=1\linewidth]{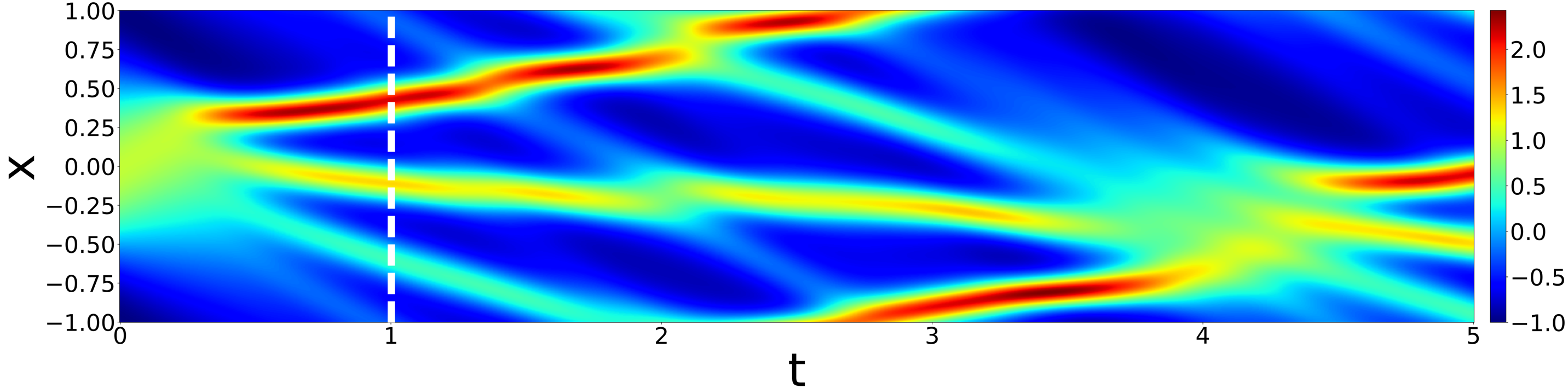}
  \caption{\small Exact solution of Korteweg–de Vries delimiting the respective $T = [0, 1]$ and $T = [0, 5]$ problems.}
  \label{fig:exact_kdv}
\end{figure}

Let us consider the following Korteweg–de Vries (KdV) problem
\begin{linenomath}\begin{align}
& \frac{\partial u}{\partial t} + u\frac{\partial u}{\partial x} + 0.0025 \frac{\partial^3 u}{\partial x^3} = 0, \; (t, x)\in [0, T] \times [-1, 1] \label{eq:kdv} 
\end{align}\end{linenomath}
subject to periodic boundary conditions and an initial condition $u(0,x) =  cos(\pi x)$. The exact solution for a short and long-time domain is shown in Figure \ref{fig:exact_kdv}.

\begin{table}[H]
	\centering
	\captionsetup{width=1\linewidth}
    \caption{\small Table of $L_2$ relative error and training time for a combination of stacked-decomposition and window-sweeping. A change in loss tolerance of $10^{-7}$ is used for all methods, with the condition in the combined form that w-s must finish propagating before s-d propagates, ensuring each subdomain is sufficiently trained given the equivalent tolerances on both methods. $^a$(bc-set = on, null-set = on), $\dag$(width = dt = $0.02$), $\ddag$(width = dt = $0.1$).}
	\begin{tabular}[c]{l | c | c}
	\toprule
        Model settings & Relative $L_2$ Error & Training time \\
        \hline
        \multicolumn{3}{c}{\multirow{2}{*}{$T\in [0, 1]$}} \\
        \multicolumn{3}{c}{} \\
        \hline
        PINN & $5.40 \times 10^{-2}$ & 2,030 \\
        s-d PINN (n = 10, dS = 1, ic = $C^p$) + FT & $1.43 \times 10^{-2}$ & 780 \\
        w-s PINN (kernel = uniform, width = dt = $0.1$)$^a$ & $1.84 \times 10^{-2}$ & 1,287 \\
        s-d + w-s$\dag$ PINN  & $2.37 \times 10^{-2}$ & 1,806 \\
        \hline
        \multicolumn{3}{c}{\multirow{2}{*}{$T\in [0, 5]$}} \\
        \multicolumn{3}{c}{} \\
        \hline
         PINN  & $9.85 \times 10^{-1}$ & 15,224 \\
         s-d PINN (n = 10, dS = 1, ic = $C^p$) + FT & $1.84 \times 10^{-1}$ & 3,566 \\
         w-s PINN (kernel = uniform, width = dt = $0.5$)$^a$ & $8.12 \times 10^{-2}$ & 16,262 \\
         s-d + w-s$\ddag$ PINN  & $5.15 \times 10^{-2}$ & 7,493 \\
	\bottomrule
	\end{tabular}
    \label{tb:kdv}
\end{table}

In Table \ref{tb:kdv}, the results for an instance of stacked-decomposition and window-sweeping are reported separately and in conjunction with one another. We observe that unmodified PINNs train well for the short time domain but encounter training difficulties, shown in \ref{ssec:kdv_long-appendix}. Although the baseline PINN trains for $T = [0, 1]$, an improvement in accuracy is still achieved from s-d and w-s PINNs with well-performing settings. We also note reduced training time in all configurations over unmodified PINNs. However, we do not observe any benefit in the combination of s-d + w-s for this domain, likely due to the lower accuracy bound for the network size and tolerances already being achieved by the methods separately. An increase in training time, therefore, follows as there is more ``overhead'' cost by using smaller window-sweeping time steps inside of stacked-decomposition subnetworks. 

\begin{figure}[H]
\centering
\includegraphics[width=1\linewidth]{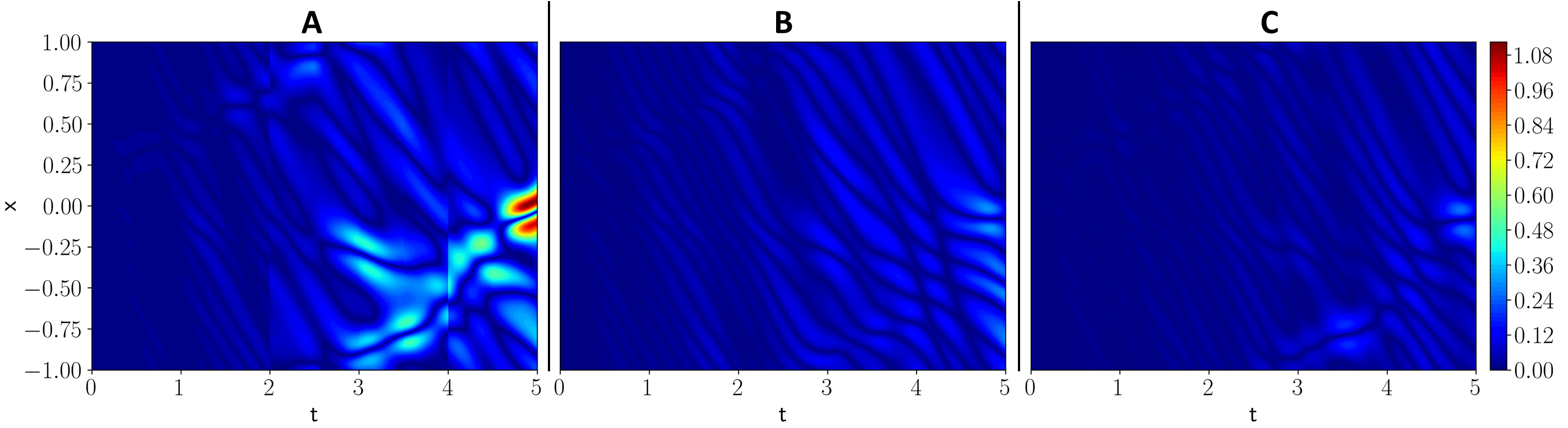}
  \caption{\small Point-wise error for $T = [0, 5]$ reported in Table \ref{tb:kdv} (A) s-d PINN (B) w-s PINN (C) s-d + w-s$\ddag$ PINN }
  \label{fig:kdv_long_pwError}
\end{figure}

In contrast, for the more difficult long-time problem, the methods on their own struggle to solve the problem well, along with unmodified PINNs. While s-d on this large domain is fast, the accuracy is poor, and w-s alone takes longer to train due to the change in loss tolerance. However, combining both yields an increase in accuracy while keeping the training time low relative to an unmodified PINN. For the w-s PINN alone, since the width and dt are large, the L-BFGS optimizer is likely to fail and cause NaNs. This is similar to why Adam is done at early training for any PINN before L-BFGS; if the domain change is too large, the optimizer is unstable. In this case, to keep the steps taken to a consistent 10, the dt = 0.5 is too large and causes optimization issues. Therefore, we perform 500 Adam iterations every time the window-sweeping scheme is propagated to ensure the stability of L-BFGS optimizer. This additional step also adds training time to the method. This is not necessary or performed for the w-s$\ddag$ setting. Figure \ref{fig:kdv_long_pwError} shows the point-wise error of the three methods used in the long-time problem. Aside from overall accuracy differences that can be inferred from the reported table values, all methods yield the highest errors at the latest time. This shows how important strongly respecting causality is, as any early deviation will lead to greater deviations later in time, regardless of the method. As the domain of a problem or the number of collocation points is increases, our framework yields greater improvements. 

\section{Summary}
\label{sec:conclusion}
We have introduced a unified framework to describe existing and new causality-enforcing PINN methods. We have showcased examples in which PINNs and their temporal decompositions can struggle to train well without modification and how settings under the proposed framework overcome these issues. Additionally, we introduce adaptive propagation strategies based on a change in loss tolerance, compared to previous versions of the methods, which use fixed optimization iterations. We achieve a reduction in training time and therefore improve scalability over an unmodified PINN on problems without training challenges. We also investigate many nuanced model decisions, such as the transferring layer parameters or enforcing boundary conditions, among others, to help guide decision-making. In future work, we will consider second-order in time problems such as the wave and Boussinesq equations, which have separate considerations when decomposing. Additionally, we will look to adapt our framework to second-order problems with only zeroth-order information at the initial and final time. This contrasts the standard setup of zero and first-order information at the initial condition. This case poses unique information propagation considerations as the standard causality approach to move forward in time would not apply. 

\vspace{0.2in}
\noindent {\bf Acknowledgements:}
This work was funded under AFOSR MURI FA9550-20-1-0358.

\appendix 
\section{Symbols and Notations}
\label{sec:appendix}

\captionof{table}{\small Symbols and Notations} 
\begin{table}[H]
	\centering
	\captionsetup{width=1\linewidth}
	\begin{tabular}[c]{l l}
    \hline
    $u(\cdot)$ & PDE solution \\
    $\boldsymbol{\theta}$ & PINN learnable parameters \\
    $u_{\boldsymbol{\theta}}(\cdot)$ & PINN PDE prediction \\
    $\mathcal{R}(\cdot)$ & PDE residual \\ 
    s-d & Stacked-decomposition \\
    w-s & Window-sweeping \\
    $N_{i}$ & Number of interface points \\
    $N_{r}$ & Number of residual collocation points \\
    $N_{ic}$ & Number of initial condition points \\
    $N_{b}$ & Number of boundary points \\
    M & Order of Fourier feature encoding \\
    $dS$ & Number of sub-networks training at once \\
    $n$ & Total number of time-slabs \\
    $\Omega$ & Spatial domain of interest \\
    $T$ & Temporal domain of interest \\
    $\mathbf{x}$ & Spatial value, $\mathbf{x} \in \Omega$ \\
    $t$  & Temporal value, $t \in T$ \\
    FT & Fine tuning \\
    TL & Transfer learning \\
    BC & Boundary Conditions \\
    bc & Backward-compatibility \\
    IC & Initial condition \\
    ic & Interface condition \\
    \hline
	\end{tabular}
\end{table}

\section{Auxiliary Results}
\label{sec:auxiliary}

\subsection{Convection: Fine Tuning vs. Transfer Learning}
\label{sec:ft_vs_tl_appendix}
\begin{figure}[H]
\centering
\includegraphics[width=\textwidth]{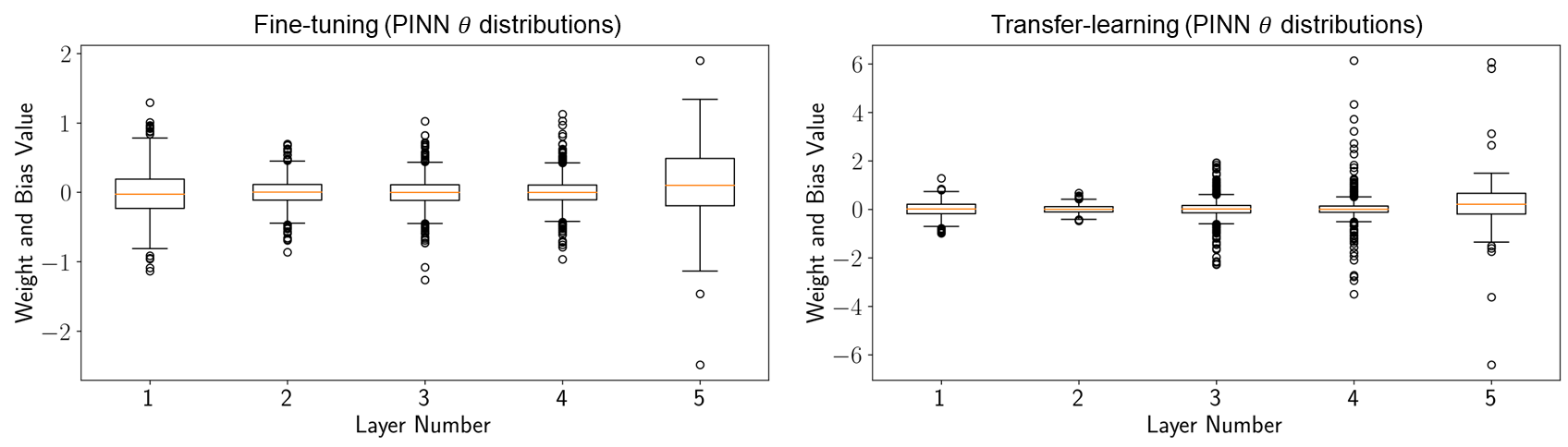}
  \caption{\small s-d PINN (n = 10, dS = 1, ic = $C^p$) learnable parameter distributions at the end of training for the final time-slab for results reported in Table \ref{tb:convec}.}
  \label{fig:Ft-vs-TR}
\end{figure}

Figure \ref{fig:Ft-vs-TR} shows the learnable parameter distributions of each layer in the final time-slab network for the results reported in Table \ref{tb:convec}. The respective models in the Table are s-d PINN (n = 10, dS = 1, ic = $C^p$) + FT and s-d PINN (n = 10, dS = 1, ic = $C^p$) + TL. We observe that when we freeze the first two layers in the network during transfer learning, the final three layers must over-adjust to compensate for the reduced expressively of the network. This can be seen in the distribution plots as the model parameter using fine tuning stays around $[-1, 1]$ while the transfer learning increases to $[-4, 4]$. In turn, this leads to longer training times for transfer learning compared to fine tuning as the model with greater expressivity more quickly converges to the solution. In contrast, the model with frozen layers must go to more extreme parameter values to satisfy the solution. Note that this observation is only in regard to temporal decomposition with PINNs and in no way is commenting on the trade-off between fine tuning and transfer learning for other applications.

\subsection{Allen-Cahn: Training Dynamics of Stacked-Decomposition \& Window-Sweeping}
\label{sec:dynamics_sd_vs_ws_appendix}

\begin{figure}[H]
\centering
\includegraphics[width=0.85\textwidth]{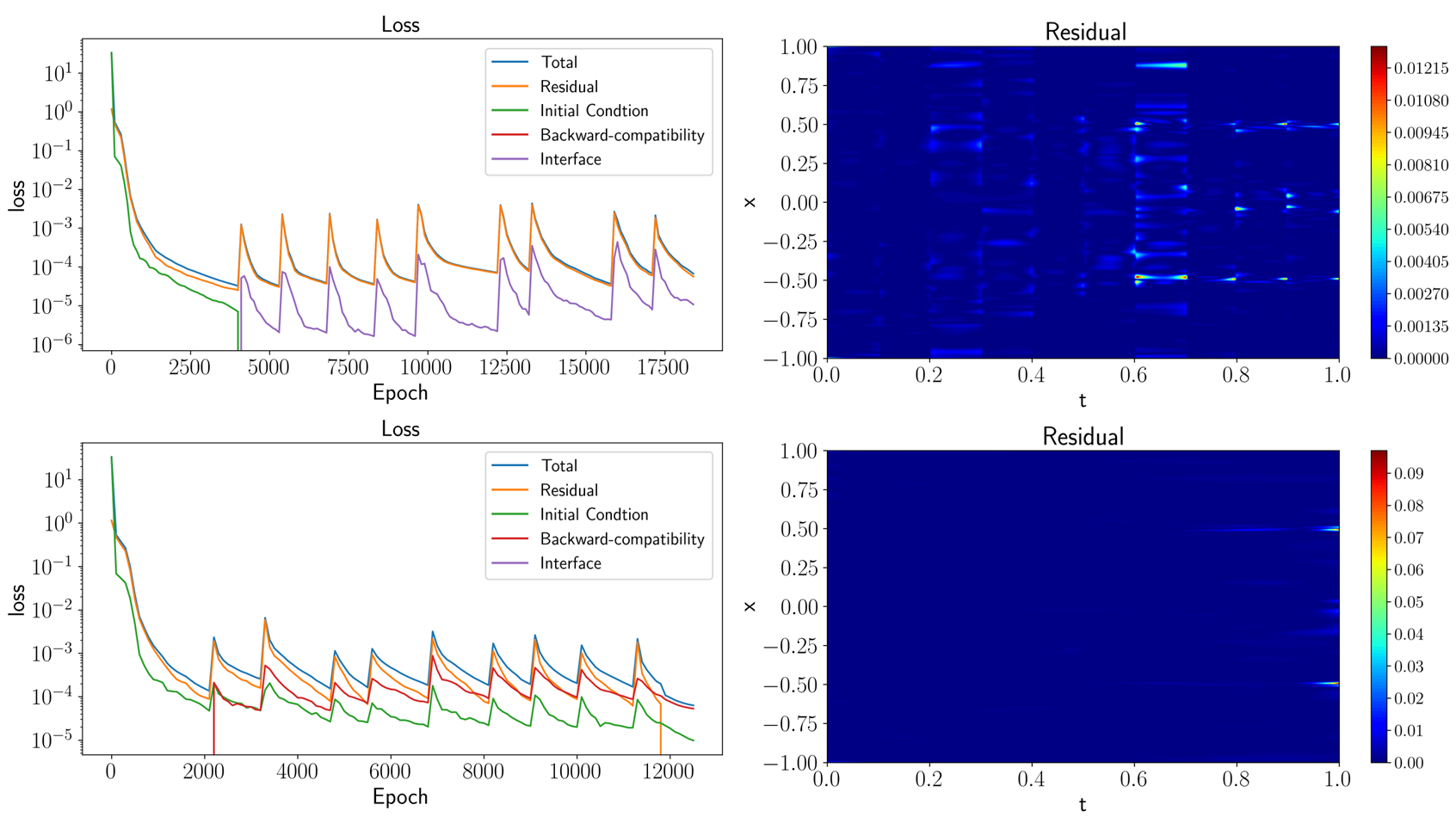}
  \caption{\small Plots of loss as a function of training epochs and the full domain PDE residual at the end of training for results reported in Table \ref{tb:allen-cahn} (Top) s-d PINN (n = 10, dS = 1, ic = $C^p$) + FT on the Allen-Cahn problem. (Bottom) w-s PINN (kernel = ``uniform'', width = dt = $0.1$)$^c$ on the Allen-Cahn problem.}
  \label{fig:sd_vs_ws}
\end{figure}

In Figure \ref{fig:sd_vs_ws}, the training dynamics are reported for stacked-decomposition and window-sweeping using settings that recover models of time-marching with fine tuning and bc-PINNs, respectively. The loss in both jumps as either a new subdomain or subnetwork is added in time-marching or as the residual subdomain moves forward, leaving the prior subdomain to be considered backward compatible in bc-PINNs. The differences are that for stacked-decomposition, unless n = dS, the initial conditions will eventually not be considered in the loss minimization. The information is purely stored and propagated through later-in-time subdomains and interfaces. With respect to window-sweeping, the initial condition will always be included during optimization. Conversely, backward-compatibility, if used, does not occur until after the initial training near the starting time, as seen in the plots. Since only one network is used in window-sweeping, the  residual and, therefore, the prediction will be continuous, whereas stacked-decomposition will have visible discontinuities at the interfaces. However, multiple networks have more expressivity than a single one as long as training challenges\cmnt{failure modes} do not occur and interfaces are well respected, leading to smaller residuals. This is particularly true at the end of time, as the residuals here contribute minimally to a global network but significantly to a local one. 

\label{ssec:loss_convergence-appendix}

\subsection{Allen-Cahn: Causal Weights}
\label{sec:causal_weights_appendix}
A loss tolerance was used to propagate all stacked-decomposition and window-sweeping methods that eliminate the fixed epoch conditions and vastly reduce computational cost. To fairly compare accuracies and training times, the loss tolerance is consistent between settings. A value of $10^{-7}$ is used for Allen-Cahn, for which a lower tolerance increases training time with no improvement to accuracy. A higher tolerance decreases accuracy, since this trade-off is discussed throughout the manuscript. We find that for higher causality parameters such as $\epsilon = (10, 100)$ described in the original paper, the change in loss reaches below $10^{-8}$ within a few hundred iterations. For smaller values, the method does not strongly enforce causality enough to overcome the training challenge. This sensitivity is addressed in the original paper by using a cascading $\epsilon$ with increasing steepness. In effect, this sweeps across the domain five times instead of once, which is not in the scope of our study, although our window-sweeping method can be used for multiple sweeps in the same way. 

Therefore, we provide self-contained results for this setting, so the reported values are not misinterpreted as advocating for or against a setting. The main contribution of the paper is to provide a unified framework in which many methods can be described, improve scalability, and generally overcome unmodified PINNs training challenges. To this end, the change in loss tolerance is removed, and a termination condition of $min_{i} w_{i} > \delta = 0.95$ is used. 

Although causal weights appear continuous, due to its implementation in \cite{wang2022respecting}, the scheme is also broken up into time snapshots like any other kernel in our window-sweeping method. This is due to the loss being formed by the mean of the mean squared error for each snapshot, unlike the standard PINN residual loss, which is the mean squared error of all points in the domain. Therefore, this formulation acts differently than the weight masks we employ in the linear and error function kernels of window-sweeping, where all the points are considered separately.

For non-grid sampling, we have attempted to run the original implementation; however, when weighting without snapshots, spatial correlation is broken, making the method fail. Therefore, we adapt the method to non-grid sampling, in this case, Latin Hypercube sampling (LHS), by treating it similarly to a grid in terms of the algorithm. Given a $100 \times 100$ grid on $T = [0,1]$, a sequence of 100 weights is generated, representing an equidistant sampling of 100 spatial points at every 0.01 increment in time. For LHS, we simply order the set of 10,000 points in time and separate them into 100-point sets of size 100 in time. This gives a similar weighting scheme to grid sampling since the mean over each of the 100 weights is used, whereas before, there were individual weights for each point. This modification is not restricted to grids and still has spatial correlation if the sampling is dense enough. 

In Table \ref{tb:allen-cahn-causal}, window-sweeping with causal weights kernel is used to solve the Allen-Cahn problem for a single pass of $\epsilon$ = 10. The modification made to alleviate the grid sampling restriction has not had any adverse effect on the method's performance. Additionally, by utilizing the null-set segmentation, which has been applied by only adding future sets (out of 100) when $min_{i} w_{i} > 0.05$, we have reduced the training time by not predicting the residual of points with negligible weights later in time. This can be extended to using the bc-set segmentation to further improve training time, as shown for the other window-sweeping kernels. Under these settings, we do not achieve the $10^{-3}$ relative $L_2$ errors reported in the original paper due to several factors. First, we use a less restrictive termination condition on $\delta$ such that the training time is in the same realm as the other kernels, which use the change in loss tolerance. We find that the difference in training time between $\delta = 0.95$ and $0.99$ is great. The success, in terms of accuracy, not cost, of the causal weights reported for Allen-Cahn in \cite{wang2022respecting} is likely largely due to the 10-100$\times$ increase in iterations 10-100$\times$ increase in network parameters $\theta$ as well as other modifications. All window-sweeping kernels reported achieving comparable results under the setting chosen. We do not make any assertion as to which method performs the best in the extreme training limit in terms of accuracy or cost, as that is not within the scope of this study.

\begin{table}[H]
	\centering
	\captionsetup{width=1\linewidth}
    \caption{\small Table of $L_2$ relative error and training time for different window-sweeping settings. All methods use M = 10 unless otherwise stated. Note that window-sweeping is abbreviated w-s. $^a$(bc-set = off, null-set = off), $^b$(bc-set = off, null-set = on)} \label{tb:allen-cahn-causal}
	\begin{tabular}[c]{l | c | c}
	\toprule
        Model settings & Relative $L_2$ Error & Training time \\
        \hline
        w-s PINN (kernel = causal weights, $\epsilon$ = 10)$^a$ + Grid sample & $3.72 \times 10^{-2}$ & 1,495 \\
        w-s PINN (kernel = causal weights, $\epsilon$ = 10)$^a$  & $3.37 \times 10^{-2}$ & 1,452 \\
        w-s PINN (kernel = causal weights, $\epsilon$ = 10)$^b$  & $4.03 \times 10^{-2}$ & 967 \\
	\bottomrule
	\end{tabular}
\end{table}

\subsection{KdV (long-time) PINN Prediction}
\label{ssec:kdv_long-appendix}

\begin{figure}[H]
\centering
\includegraphics[width=\textwidth]{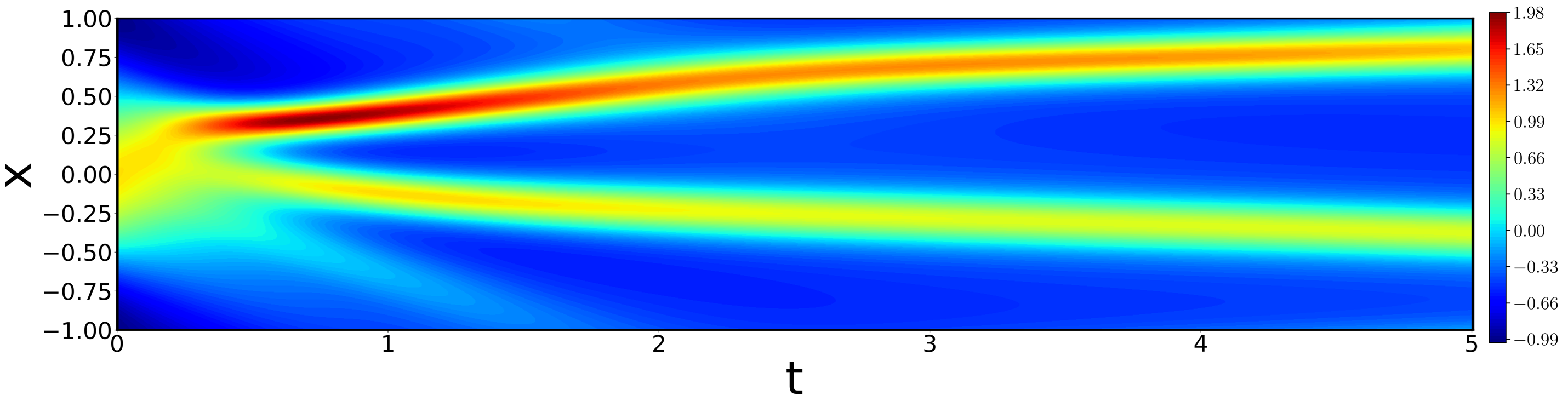}
  \caption{\small PINN prediction for the long-time KdV problem in Section \ref{ssec:kdv}, representative of the incorrect propagation training challenge\cmnt{failure}.}
  \label{fig:kdv_long_PINN}
\end{figure}

In Figure \ref{fig:kdv_long_PINN}, while the PINN solves the KdV problem for $T = [0,1]$, it fails for $T = [0,5]$. Interestingly, the issue\cmnt{failure mode} is not one of the zero-solution as in the long-time Convection problem but is in fact the incorrect propagation challenge \cmnt{failure}such as in the Allen-Cahn problem. This is likely due to the traveling wave in the convection problem being extended due to the periodic conditions. While that feature is not present in the KdV problem, the increased training difficulty of a larger temporal domain manifests itself with incorrect information propagation instead. 

\section{Fourier Feature encoding ($\mathcal{C}^{\infty}$ periodic conditions)}
\label{sec:fourier_appendix}
Following the work from \cite{dong2021method, wang2022respecting}, we can exactly enforce $\mathcal{C}^{\infty}$ periodic boundary conditions by applying a Fourier feature encoding to the spatial input of the network. The spatial encoding is
\begin{align}
v(x) = \{ 1, cos(\omega x), sin (\omega x), ..., cos(M\omega x), sin (M\omega x)\}
\end{align}
where $\omega = \frac{2 \pi}{L}$, $L = x_{max} - x_{min}$, and M is a non-negative integer representing the sinusoidal frequency of the input. A higher M leads to even higher frequency components in the output after passing through nonlinear activation functions, which may be helpful in PDE problems with high-frequency solution components such as the Allen-Cahn problem considered here. All choices of M are shown to be $\mathcal{C}^{\infty}$ periodic in Lemma 2.1 of \cite{dong2021method}.

\newpage
\noindent {\bf Bibliography}
\bibliography{references}

\begin{thebibliography}{10}
\expandafter\ifx\csname url\endcsname\relax
  \def\url#1{\texttt{#1}}\fi
\expandafter\ifx\csname urlprefix\endcsname\relax\def\urlprefix{URL }\fi
\expandafter\ifx\csname href\endcsname\relax
  \def\href#1#2{#2} \def\path#1{#1}\fi

\bibitem{raissi2019physics}
M.~Raissi, P.~Perdikaris, G.~E. Karniadakis, {Physics-informed neural networks:
  A deep learning framework for solving forward and inverse problems involving
  nonlinear partial differential equations}, Journal of Computational Physics
  378 (2019) 686--707.

\bibitem{mathews2021uncovering}
A.~Mathews, M.~Francisquez, J.~W. Hughes, D.~R. Hatch, B.~Zhu, B.~N. Rogers,
  {Uncovering turbulent plasma dynamics via deep learning from partial
  observations}, Physical Review E 104~(2) (2021) 025205.

\bibitem{kissas2020machine}
G.~Kissas, Y.~Yang, E.~Hwuang, W.~R. Witschey, J.~A. Detre, P.~Perdikaris,
  {Machine learning in cardiovascular flows modeling: Predicting arterial blood
  pressure from non-invasive 4D flow MRI data using physics-informed neural
  networks}, Computer Methods in Applied Mechanics and Engineering 358 (2020)
  112623.

\bibitem{10.1371/journal.pcbi.1007575}
A.~Yazdani, L.~Lu, M.~Raissi, G.~E. Karniadakis,
  \href{https://doi.org/10.1371/journal.pcbi.1007575}{{Systems biology informed
  deep learning for inferring parameters and hidden dynamics}}, PLOS
  Computational Biology 16 (2020) 1--19.
\newblock \href {http://dx.doi.org/10.1371/journal.pcbi.1007575}
  {\path{doi:10.1371/journal.pcbi.1007575}}.
\newline\urlprefix\url{https://doi.org/10.1371/journal.pcbi.1007575}

\bibitem{WANG2021109914}
S.~Wang, P.~Perdikaris,
  \href{https://www.sciencedirect.com/science/article/pii/S0021999120306884}{{Deep
  learning of free boundary and Stefan problems}}, Journal of Computational
  Physics 428 (2021) 109914.
\newblock \href {http://dx.doi.org/https://doi.org/10.1016/j.jcp.2020.109914}
  {\path{doi:https://doi.org/10.1016/j.jcp.2020.109914}}.
\newline\urlprefix\url{https://www.sciencedirect.com/science/article/pii/S0021999120306884}

\bibitem{shukla2020physics}
K.~Shukla, P.~C. Di~Leoni, J.~Blackshire, D.~Sparkman, G.~E. Karniadakis,
  {Physics-informed neural network for ultrasound nondestructive quantification
  of surface breaking cracks}, Journal of Nondestructive Evaluation 39 (2020)
  1--20.

\bibitem{Chen:20}
Y.~Chen, L.~Lu, G.~E. Karniadakis, L.~D. Negro,
  \href{https://opg.optica.org/oe/abstract.cfm?URI=oe-28-8-11618}{{Physics-informed
  neural networks for inverse problems in nano-optics and metamaterials}}, Opt.
  Express 28~(8) (2020) 11618--11633.
\newblock \href {http://dx.doi.org/10.1364/OE.384875}
  {\path{doi:10.1364/OE.384875}}.
\newline\urlprefix\url{https://opg.optica.org/oe/abstract.cfm?URI=oe-28-8-11618}

\bibitem{10.3389/fphy.2020.00042}
F.~Sahli~Costabal, Y.~Yang, P.~Perdikaris, D.~E. Hurtado, E.~Kuhl,
  \href{https://www.frontiersin.org/articles/10.3389/fphy.2020.00042}{{Physics-Informed
  Neural Networks for Cardiac Activation Mapping}}, Frontiers in Physics 8.
\newblock \href {http://dx.doi.org/10.3389/fphy.2020.00042}
  {\path{doi:10.3389/fphy.2020.00042}}.
\newline\urlprefix\url{https://www.frontiersin.org/articles/10.3389/fphy.2020.00042}

\bibitem{cuomo2022scientific}
S.~Cuomo, V.~S. Di~Cola, F.~Giampaolo, G.~Rozza, M.~Raissi, F.~Piccialli,
  Scientific machine learning through physics--informed neural networks: Where
  we are and what’s next, Journal of Scientific Computing 92~(3) (2022) 88.

\bibitem{hao2022physics}
Z.~Hao, S.~Liu, Y.~Zhang, C.~Ying, Y.~Feng, H.~Su, J.~Zhu, Physics-informed
  machine learning: A survey on problems, methods and applications, arXiv
  preprint arXiv:2211.08064.

\bibitem{krishnapriyan2021characterizing}
A.~Krishnapriyan, A.~Gholami, S.~Zhe, R.~Kirby, M.~W. Mahoney, {Characterizing
  possible failure modes in physics-informed neural networks}, Advances in
  Neural Information Processing Systems 34 (2021) 26548--26560.

\bibitem{wight2020solving}
C.~L. Wight, J.~Zhao, {Solving Allen Cahn and Cahn Hilliard equations using the
  adaptive physics informed neural networks}, arXiv preprint arXiv:2007.04542.

\bibitem{hu2021extended}
Z.~Hu, A.~D. Jagtap, G.~E. Karniadakis, K.~Kawaguchi, {When do extended
  physics-informed neural networks (XPINNs) improve generalization?}, SIAM
  Journal on Scientific Computing 44~(5) (2022) A3158--A3182.

\bibitem{mojgani2022lagrangian}
R.~Mojgani, M.~Balajewicz, P.~Hassanzadeh,
  \href{https://www.sciencedirect.com/science/article/pii/S0045782522007666}{{Kolmogorov
  n–width and Lagrangian physics-informed neural networks: A
  causality-conforming manifold for convection-dominated PDEs}}, Computer
  Methods in Applied Mechanics and Engineering 404 (2023) 115810.
\newblock \href {http://dx.doi.org/https://doi.org/10.1016/j.cma.2022.115810}
  {\path{doi:https://doi.org/10.1016/j.cma.2022.115810}}.
\newline\urlprefix\url{https://www.sciencedirect.com/science/article/pii/S0045782522007666}

\bibitem{jagtap2022deep}
A.~D. Jagtap, D.~Mitsotakis, G.~E. Karniadakis, {Deep learning of inverse water
  waves problems using multi-fidelity data: Application to Serre--Green--Naghdi
  equations}, Ocean Engineering 248 (2022) 110775.

\bibitem{jagtap2022physics}
A.~D. Jagtap, Z.~Mao, N.~Adams, G.~E. Karniadakis, {Physics-informed neural
  networks for inverse problems in supersonic flows}, Journal of Computational
  Physics 466 (2022) 111402.

\bibitem{9664609}
K.~Shukla, A.~D. Jagtap, J.~L. Blackshire, D.~Sparkman, G.~Em~Karniadakis, {A
  Physics-Informed Neural Network for Quantifying the Microstructural
  Properties of Polycrystalline Nickel Using Ultrasound Data: A promising
  approach for solving inverse problems}, IEEE Signal Processing Magazine
  39~(1) (2022) 68--77.
\newblock \href {http://dx.doi.org/10.1109/MSP.2021.3118904}
  {\path{doi:10.1109/MSP.2021.3118904}}.

\bibitem{inverse_groundwater}
I.~Depina, S.~Jain, S.~M. Valsson, H.~Gotovac, {Application of physics-informed
  neural networks to inverse problems in unsaturated groundwater flow},
  Georisk: Assessment and Management of Risk for Engineered Systems and
  Geohazards 16~(1) (2022) 21--36.
\newblock \href {http://dx.doi.org/10.1080/17499518.2021.1971251}
  {\path{doi:10.1080/17499518.2021.1971251}}.

\bibitem{chen2020physics}
Y.~Chen, L.~Lu, G.~E. Karniadakis, L.~Dal~Negro, {Physics-informed neural
  networks for inverse problems in nano-optics and metamaterials}, Optics
  express 28~(8) (2020) 11618--11633.

\bibitem{mishra2020estimates}
S.~Mishra, R.~Molinaro, {Estimates on the generalization error of Physics
  Informed Neural Networks (PINNs) for approximating a class of inverse
  problems for PDEs}, arXiv preprint arXiv:2007.01138.

\bibitem{thakur2023temporal}
S.~Thakur, M.~Raissi, H.~Mitra, A.~Ardekani, Temporal consistency loss for
  physics-informed neural networks, arXiv preprint arXiv:2301.13262.

\bibitem{lu2021deepxde}
L.~Lu, X.~Meng, Z.~Mao, G.~E. Karniadakis, {DeepXDE: A deep learning library
  for solving differential equations}, SIAM Review 63~(1) (2021) 208--228.

\bibitem{daw2022rethinking}
A.~Daw, J.~Bu, S.~Wang, P.~Perdikaris, A.~Karpatne, {Rethinking the Importance
  of Sampling in Physics-informed Neural Networks}, arXiv preprint
  arXiv:2207.02338.

\bibitem{subramanian2022adaptive}
S.~Subramanian, R.~M. Kirby, M.~W. Mahoney, A.~Gholami, {Adaptive
  Self-supervision Algorithms for Physics-informed Neural Networks}, arXiv
  preprint arXiv:2207.04084.

\bibitem{wang2022respecting}
S.~Wang, S.~Sankaran, P.~Perdikaris, {Respecting causality is all you need for
  training physics-informed neural networks}, arXiv preprint arXiv:2203.07404.

\bibitem{mcclenny2020self}
L.~McClenny, U.~Braga-Neto, {Self-adaptive physics-informed neural networks
  using a soft attention mechanism}, arXiv preprint arXiv:2009.04544.

\bibitem{jagtap2020adaptive}
A.~D. Jagtap, K.~Kawaguchi, G.~E. Karniadakis, {Adaptive activation functions
  accelerate convergence in deep and physics-informed neural networks}, Journal
  of Computational Physics 404 (2020) 109136.

\bibitem{jagtap2022important}
A.~D. Jagtap, G.~E. Karniadakis, {How important are activation functions in
  regression and classification? A survey, performance comparison, and future
  directions}, arXiv preprint arXiv:2209.02681 (2022).

\bibitem{yu2022gradient}
J.~Yu, L.~Lu, X.~Meng, G.~E. Karniadakis, {Gradient-enhanced physics-informed
  neural networks for forward and inverse PDE problems}, Computer Methods in
  Applied Mechanics and Engineering 393 (2022) 114823.

\bibitem{JagtapK}
A.~D. Jagtap, G.~E. Karniadakis, {Extended Physics-Informed Neural Networks
  (XPINNs): A Generalized Space-Time Domain Decomposition Based Deep Learning
  Framework for Nonlinear Partial Differential Equations}, Communications in
  Computational Physics 28 (2020) 2002--2041.
\newblock \href {http://dx.doi.org/10.4208/cicp.OA-2020-0164}
  {\path{doi:10.4208/cicp.OA-2020-0164}}.

\bibitem{JAGTAP2020113028}
A.~D. Jagtap, E.~Kharazmi, G.~E. Karniadakis,
  \href{https://www.sciencedirect.com/science/article/pii/S0045782520302127}{{Conservative
  physics-informed neural networks on discrete domains for conservation laws:
  Applications to forward and inverse problems}}, Computer Methods in Applied
  Mechanics and Engineering 365 (2020) 113028.
\newblock \href {http://dx.doi.org/https://doi.org/10.1016/j.cma.2020.113028}
  {\path{doi:https://doi.org/10.1016/j.cma.2020.113028}}.
\newline\urlprefix\url{https://www.sciencedirect.com/science/article/pii/S0045782520302127}

\bibitem{meng2020ppinn}
X.~Meng, Z.~Li, D.~Zhang, G.~E. Karniadakis, {PPINN: Parareal physics-informed
  neural network for time-dependent PDEs}, Computer Methods in Applied
  Mechanics and Engineering 370 (2020) 113250.

\bibitem{hu2022augmented}
Z.~Hu, A.~D. Jagtap, G.~E. Karniadakis, K.~Kawaguchi, {Augmented
  Physics-Informed Neural Networks (APINNs): A gating network-based soft domain
  decomposition methodology}, arXiv preprint arXiv:2211.08939 (2022).

\bibitem{braga2022characteristics}
U.~Braga-Neto, {Characteristics-Informed Neural Networks for Forward and
  Inverse Hyperbolic Problems}, arXiv preprint arXiv:2212.14012.

\bibitem{shin2020convergence}
Y.~Shin, J.~Darbon, G.~E. Karniadakis, {On the convergence of physics informed
  neural networks for linear second-order elliptic and parabolic type PDEs},
  arXiv preprint arXiv:2004.01806.

\bibitem{mishra2022estimates}
S.~Mishra, R.~Molinaro, {Estimates on the generalization error of
  physics-informed neural networks for approximating a class of inverse
  problems for PDEs}, IMA Journal of Numerical Analysis 42~(2) (2022)
  981--1022.

\bibitem{de2022error}
T.~De~Ryck, A.~D. Jagtap, S.~Mishra, {Error estimates for physics informed
  neural networks approximating the Navier-Stokes equations}, arXiv preprint
  arXiv:2203.09346 (2022).

\bibitem{wang2021understanding}
S.~Wang, Y.~Teng, P.~Perdikaris, {Understanding and mitigating gradient flow
  pathologies in physics-informed neural networks}, SIAM Journal on Scientific
  Computing 43~(5) (2021) A3055--A3081.

\bibitem{wang2021eigenvector}
S.~Wang, H.~Wang, P.~Perdikaris, {On the eigenvector bias of Fourier feature
  networks: From regression to solving multi-scale PDEs with physics-informed
  neural networks}, Computer Methods in Applied Mechanics and Engineering 384
  (2021) 113938.

\bibitem{MATTEY2022114474}
R.~Mattey, S.~Ghosh,
  \href{https://www.sciencedirect.com/science/article/pii/S0045782521006939}{{A
  novel sequential method to train physics informed neural networks for Allen
  Cahn and Cahn Hilliard equations}}, Computer Methods in Applied Mechanics and
  Engineering 390 (2022) 114474.
\newblock \href {http://dx.doi.org/https://doi.org/10.1016/j.cma.2021.114474}
  {\path{doi:https://doi.org/10.1016/j.cma.2021.114474}}.
\newline\urlprefix\url{https://www.sciencedirect.com/science/article/pii/S0045782521006939}

\bibitem{raissi2017physicsI}
M.~Raissi, P.~Perdikaris, G.~E. Karniadakis, {Physics Informed Deep Learning
  (Part I): Data-driven Solutions of Nonlinear Partial Differential Equations},
  arXiv preprint arXiv:1711.10561.

\bibitem{raissi2017physicsII}
M.~Raissi, P.~Perdikaris, G.~E. Karniadakis, {Physics Informed Deep Learning
  (Part II): Data-driven Discovery of Nonlinear Partial Differential
  Equations}, arXiv preprint arXiv:1711.10566.

\bibitem{dong2021method}
S.~Dong, N.~Ni, {A method for representing periodic functions and enforcing
  exactly periodic boundary conditions with deep neural networks}, Journal of
  Computational Physics 435 (2021) 110242.

\bibitem{wang2022and}
S.~Wang, X.~Yu, P.~Perdikaris, {When and why PINNs fail to train: A neural
  tangent kernel perspective}, Journal of Computational Physics 449 (2022)
  110768.

\bibitem{DeepLearning}
I.~Goodfellow, Y.~Bengio, A.~Courville, {Deep Learning}, The MIT Press,
  Cambridge, MA, USA, 2016.

\bibitem{kingma2014adam}
D.~P. Kingma, J.~Ba, {Adam: A method for stochastic optimization}, arXiv
  preprint arXiv:1412.6980.

\bibitem{liu1989limited}
D.~C. Liu, J.~Nocedal, {On the limited memory BFGS method for large scale
  optimization}, Mathematical programming 45~(1-3) (1989) 503--528.

\bibitem{bihlo2022physics}
A.~Bihlo, R.~O. Popovych, {Physics-informed neural networks for the
  shallow-water equations on the sphere}, Journal of Computational Physics 456
  (2022) 111024.

\bibitem{shukla2021parallel}
K.~Shukla, A.~D. Jagtap, G.~E. Karniadakis, {Parallel physics-informed neural
  networks via domain decomposition}, Journal of Computational Physics 447
  (2021) 110683.

\bibitem{li2022meta}
S.~Li, M.~Penwarden, R.~M. Kirby, S.~Zhe, Meta learning of interface conditions
  for multi-domain physics-informed neural networks, arXiv preprint
  arXiv:2210.12669.

\bibitem{Driscoll2014}
T.~A. Driscoll, N.~Hale, L.~N. Trefethen,
  \href{http://www.chebfun.org/docs/guide/}{{Chebfun Guide}}, Pafnuty
  Publications, 2014.
\newline\urlprefix\url{http://www.chebfun.org/docs/guide/}

\bibitem{COX2002430}
S.~Cox, P.~Matthews, {Exponential Time Differencing for Stiff Systems}, Journal
  of Computational Physics 176~(2) (2002) 430--455.
\newblock \href {http://dx.doi.org/https://doi.org/10.1006/jcph.2002.6995}
  {\path{doi:https://doi.org/10.1006/jcph.2002.6995}}.

\end{thebibliography}

\end{document}